\documentclass[11pt]{article}
\usepackage{amssym}
\begin{document}

\author{Vladimir K. Petrov}
\title{Scaling in a toy model of gluodynamics at finite temperatures}
\date{\textit{N. N. Bogolyubov Institute for Theoretical Physics}\\
\textit{\ National Academy of Sciences of Ukraine}\\
\textit{\ 252143 Kiev, Ukraine. 12.15.1997}}
\maketitle

\begin{abstract}
In the limit of $\xi \simeq a_\sigma /a_\tau \rightarrow \infty $ the
gluodynamics without the magnetic part of action ( $S_M\sim 1/\xi $) is
considered as a self-contained model. The model is studied analytically in
the continuum limit on an extremely large lattice ( $N_\tau \rightarrow
\infty $ ). Scaling conditions for critical temperature and string tension
are considered. The model shows trivial ( $g^2\sim a_\tau $) asymptotic
freedom in the case of continuous gauge groups and nontrivial one ( $g^2\sim
1/\ln 1/a_\tau $) for discrete groups.

\endabstract
\pagebreak[4]
\end{abstract}

\section{Introduction. MC experiment and some problems of renormalisation
procedure in LGT}

Since the confirmation of the approach to the continuum limit is of
particular importance in LGT, the systematic scaling analysis at larger $%
\beta =\frac{2N}{g^2}$ seams inevitable . In lattice QCD the
non-perturbative aspects of the theory are of primary interest, but the
renormalisation-group techniques are added , as a rule, in a perturbative
way. Feasible enough description of the running coupling\texttt{\ }behavior
by field theory methods allows to establish a direct connection with the
standard formalism of perturbation theory and the renormalisation procedure
in lattice gauge theory (LGT). Now the Callan-Symanzik beta\ function 
\begin{equation}
\beta _{CS}\left( g\right) =-a\frac{\partial g}{\partial a}%
=-b_0g^3-b_1g^5-b_2g_0^7-...  \label{bet}
\end{equation}
which leads to 
\begin{equation}
\quad g^{-2}=b_0\ln \frac 1{a\Lambda _L}+\frac{b_1}{b_0}\ln \ln \frac
1{a\Lambda _L}+...  \label{G}
\end{equation}
with 
\begin{equation}
b_0=\frac{11}3\frac N{16\pi ^2};\quad b_1=\frac{34}3\left( \frac N{16\pi
^2}\right) ^2  \label{ren}
\end{equation}
is computed in a perturbative theory with\texttt{\ }an accuracy even greater
than that required in LGT at present. Outside the weak coupling region,
however, the scaling behavior is described by the beta\ function, which
,generally speaking, is unreachable for the perturbation theory.

In the lattice gauge theory, the conditions, which provide asymptotic
scaling in a strong coupling regime were established long ago (see, e.g.,%
\cite{Creutz}). In the case of anisotropic lattice $a_\sigma /$ $a_\tau \neq
1$ such conditions can be written as 
\begin{equation}
\beta _{\sigma ,\tau }\simeq \bar \beta _{\sigma ,\tau }\exp \left( a_\sigma
a_{\sigma ,\tau }M_{\sigma ,\tau }^2\right) \rightarrow \bar \beta _{\sigma
,\tau }=const;\quad M_{\sigma ,\tau }=const  \label{g-loop}
\end{equation}
\begin{equation}
\beta _{\sigma ,\tau }\simeq \bar \beta _{\sigma ,\tau }\exp \left(
a_{\sigma ,\tau }m_{\sigma ,\tau }\right) \rightarrow \bar \beta _{\sigma
,\tau }^{\prime }=const;\quad m_{\sigma ,\tau }=const  \label{g-gap}
\end{equation}

Although $\left( \ref{g-gap}\right) $ and $\left( \ref{g-loop}\right) $
guarantee the scaling behavior of mass gaps $m_{\sigma ,\tau }$ and string
tensions $\alpha _{\sigma ,\tau }$ they cannot be satisfied simultaneously.
In both cases the Callan-Symanzik beta function $\beta _{CS}\simeq $\ $\beta
_{\sigma ,\tau }\ln \frac{\bar \beta _{\sigma ,\tau }}{\beta _{\sigma ,\tau }%
}$ turns to zero at the finite value of bare coupling, which scarcely can be
regarded as a 'strong' one. Moreover, since some physical quantities have
different strong coupling expansions, it is hardly possible to adjust the
couplings in such a way, that all of them will remain simultaneously
invariant under changing $\xi $ \cite{karsch}. All this makes us doubt that
the continuum limit may be realized in the strong coupling area.

Even being computed in a weak coupling area Eq. $\left( \ref{bet}\right) $
may differ in different approaches. In particular, only two leading terms in 
$\left( \ref{bet}\right) $ are universal. Such disturbances are annoying but
hardly significant, because the goal of lattice calculation is to discover
the value of some physical observable when the UV cutoff is taken very
large, so that the physics is, indeed, in the first term of $\left( \ref{bet}%
\right) $. Everything else is an artifact of calculation.

However,\ although the continuum limit ($a\rightarrow 0$ ) in asymptotically
free theories corresponds to $g\rightarrow 0$ such theories in the continuum
limit do not become pertubative \cite{hasenfratz}. For example, while
calculating $R\times R$ Wilson loop expectation value in perturbation
theory, the correction to the leading term typically has the form 
\begin{equation}
g^2\ln R=g^2\ln \frac ra\rightarrow \infty ;\quad r\rightarrow \infty ,
\end{equation}
so that this correction becomes arbitrary large at large distances.

At short distances, i.e.\ for energy scales larger than the heavy-quark
mass, the physics is perturbative and described by ordinary QCD. For mass
scales much below the heavy-quark mass, the physics is essentially
non-perturbative because of confinement. Thus, although a trend in
approaching to the asymptotic scaling in 2-loop approximation is seen at $%
g^2\sim 1$, one cannot exclude that the non-perturbative effects may
seriously change Eq. $\left( \ref{bet}\right) $ at extremely small $g^2$.

At the moment, lattice simulations are the most popular way of extracting
truly non-perturbative results from quantum field theories. The lattice
calculation of a number of physical values, e.g. of the glueball mass,
string tension, ets., at different $g$ may be used to find the beta\
function, these quantities, however, reflect the long distance properties.
The exception is the Creutz ratio of Wilson loops in which the self mass
contributions are cancelled, so it is, indeed, a candidate for the study of
beta\ function. Although many interesting attempts are made to obtain the
QCD running coupling from the first principles by lattice computations,
however, two old problems \cite{HHK} are still far from an ultimate solution:
the ratios of small loops are contaminated by lattice artifacts which
results in the systematic error $\backsim \frac{2N}{g^2}$ and finite size
effects\texttt{(}if the correlation length is comparable with the lattice
size).

Detailed analysis \cite{Luscher} (see also \cite{BG}) of the recent
developments in renormalisation technique shows that an important
uncertainty arises due to the fact that the results of renormalisation
procedure depend on the details of the lattice regularization. Since so
called \textit{bare perturbation theory} is regarded as unreliable, various
recipes, based on mean-field theory or resummations of tadpole diagrams,
have been suggested to deal with this problem \cite{Parisi,LM}. However,
different prescriptions give different results and it is in any case unclear
how errors can be reliably assessed.

An interesting method to compute renormalization factors that does not rely
on bare perturbation theory has been proposed in~\cite{M}. However this
method has its own problems, the most important being that the momentum $p$\
should be significantly smaller than $1/a$\ to suppress the lattice effects,
but not too small since in this case you cannot apply the renormalized
perturbation theory with desired confidence.

A method which may provide general solution of the non-perturbative
renormalization problem through a recursive procedure~\cite{LWW} is
suggested~in \cite{L,D,J,C}. In such a treatment the scale evolution of the
renormalized parameters and operators can be studied by changing the lattice
size $L$ at fixed bare parameters. With an accuracy up to lattice effects
the running couplings on the two lattices are then related through 
\begin{equation}
g^2(2L)=\sigma (g^2(L)),  \label{rec}
\end{equation}
where $\sigma $ is an integrated form of the Callan-Symanzik beta\ function.
The recursive procedure has been used to compute the running coupling in
quenched QCD~\cite{L,D} and the running quark masses \cite{J,C}. An
important point to note is that the renormalization group invariant quark
masses $M$ are scheme-independent. Such procedure, however, is not totally
independent of the initial choice of the scaling functions $\sigma (g^2(L))$
in $\left( \ref{rec}\right) $.

Let us consider some problems of renormalisation in LGT\texttt{\ }%
immediately revealed by the MC experiment. Comprehensive studies LGT on
lattices of up to $48^3\times 16$ at finite temperature \cite{Biele} and up
to $48^3\times 56$ at zero temperature \cite{booth} show that
(asymptotic)terms violating the scaling are still strong at $\beta $ values
accessible for todays numerical simulations. These terms manifest themselves
most strikingly in the famous dip of the ``step function'', $\Delta _\beta
\equiv \beta _{CS}\left( a\right) -\beta _{CS}\left( 2a\right) $ referring
to the deconfinement temperature \cite{Biele}. This pattern is similar in
all $SU(N)$ gauge theories. The strongest sensitivity was observed in the $%
N_\tau =4$ case, where the numerical calculations close to $T_c$ are
performed with bare couplings $\beta \sim 5.6\div 5.8$, i.e.\/~in the region
of the strongest variation of the cut-off with $\beta $ (dip in the $\Delta
_\beta $) \cite{Biele}. Besides, with the 'dip' in $\Delta _\beta $ the
presence of even a zero in $\Delta _\beta $ within this region of $\beta $
may be suspected \cite{patr-seiler}. Numerical studies showed deviations
from $\left( \ref{bet}\right) $ on entering the scaling region where the
correlation length $\varsigma $ begins to grow. It is especially worth to
note, that these deviations are always of such a pattern that $\varsigma $
increases faster with $\beta $ than is predicted by $\left( \ref{bet}\right) 
$, as if the theory approaches the critical point and consequently is not
asymptotically free \cite{patr-seiler}.

Recent analyses \cite{akemi} showed that scaling violations persist in
physical quantities such as string tension and hadron masses up to $\beta
=6.8$ \cite{Uk}. Above the deconfining transition point, the matching
procedure inevitably suffers large errors. MCRG has been performed by
several groups on 16$^4$ lattices in the large $\beta $ region up to $7.2$ 
\cite{HHK,B,B2,GKPS,H-k}. However, these results were inconclusive and even
controversial. Bowler et al. \cite{B,B2} found\ sizeable deviations from
2-loop scaling. Recently Hoek has reanalyzed the same data and claimed a
very slow approach to scaling \cite{H-k}. Studies on larger lattices with
better statistics are required to clarify the scaling behavior in the high
region of $\beta $.

Unfortunately, with present computer capacity we are still far enough from
the scaling region and $a$-dependent terms are large, so we can hardly
finally decide if the perturbative beta\ function $\left( \ref{bet}\right) $
is appropriate for LGT at long distance area. However, if for the
calculation of some physical value (e.g. critical temperature $T_c$ ) one
applies the beta\ function computed within LGT for another physical value
(e.g. string tension $\sigma $) essentially better results may be expected.
Moreover, $\sigma $ itself may be used an instrument of phenomenological
renormalisation procedure. Indeed, scaling behavior for $\sqrt{\sigma }/T_c$
is excellent \cite{karsch}, \cite{bali1}. MC computations in $3$ dimensions
of ratio $\sqrt{\sigma _3}/T_c$ for the $SU(2)$ and $SU(3)$ cases \cite
{bali1}, \cite{boyd} as a matter of fact, show$\ $only minor deviations from
scaling. This, however, can be taken as an evidence in favor of the
universality of the renormalisation procedure, rather than of the
applicability of perturbative beta\ function.

Unfortunately, on the basis of todays numerical computations,it is difficult
to anticipate beta function behavior in the limit of $a\rightarrow 0,$
taking perturbative calculations as a guide\footnote{%
Warnings concerning the enhancement of nonperturbative effects as $%
a\!\rightarrow \!0$\ in the presence of power divergences have been issued 
\cite{MMS}}. Therefore in the near future, at least at this point,
analytical methods will be still needed. Though one may well believe that
deviations from scaling will disappear completely with increasing lattice
sizes, a direct check of this suggestion is rather difficult because of
limited computer capacity. In our previous article \cite{petrov} we made an
attempt to compute analytically spatial string tension $\sqrt{\sigma }$\ in $%
Z\left( 2\right) $\ and $Z\left( 3\right) $\ gluodynamics at finite
temperature in spherical model approximation. As it was shown, the spatial
string tension demonstrated scaling behavior for certain parameter area $%
b_0\approx \frac \xi {2N}\left( 1-\cos \frac{2\pi }N\right) ^{-1}$. The
numeric value of $b_0$\ with anisotropy parameter $\xi \simeq 1$ agree with
the standard value $b_0=\frac{11}{12}\frac N{4\pi ^2}$. In this article we
try to find analytically the running coupling dependence on lattice sizes
taking some simple cases and compare such a dependence for models with
continuous and discrete gauge groups. For this purpose we reconsider the
well known effective action for finite\ temperature gluodynamics \cite{green}%
, that helps us to estimate the critical temperature and running coupling
behavior in continuum limit on 'large' lattices ($N_\tau \rightarrow \infty $%
).

The most serious problem appearing in such approach is the calculation of
the contribution of the magnetic part ($S_M$) of the action . It was studied
in \cite{billo} where it had been established that pertubative corrections
slowed down the critical temperature dependance on $N_\tau $. However, there
are reasons to doubt that the magnetic part contribution can be reliably
computed perturbatively. Below we give some evidence in favor of the
suggestion that in $N_\tau \rightarrow \infty $ limit the situation can
hardly be totally cured by the computation of the finite set of perturbative
series terms.

Being unable to make such calculations nonperturbatively we consider an
'electric toy' model, where all terms of order $1/$ $\xi $ \footnote{%
Asymmertry (or anisitropy) parameter is defined as $\xi \simeq a_\sigma
/a_\tau ,$ where $a_\tau $ and $a_\sigma $ are temporal and spatial lattice
spacings.} in $QCD$ action are omitted. In particular the magnetic part of
the action in this model is absent. Although such approach can be regarded
as too rough and unrealistic, it may be interesting from mathematical point
of view, as an example of a model, where renormalisation procedure can be
performed analytically entirely within the lattice theory. In fact, very
similar approximation was partially used for the analysis of pure gauge
theories in \cite{svetitsky} where $T=\frac 1{a_\tau N_\tau }\rightarrow
\infty $ was additionally claimed. As it was mentioned above, along with $%
a_\tau \rightarrow 0$ we consider the limit of $N_\tau \rightarrow \infty $
, therefore in suggested model the finite temperatures area also becomes
accessible . Some results obtained in such toy model may be useful for
general considerations of renormalisation procedure in LGT.

\section{Effective action for finite\ temperature $Z\left( N\right) $
gluodynamics}

Here we neglect the contribution of space-like plaquettes i.e. the
chromomagnetic part of the action. Of course, one may do it in a strong
coupling approximation, however, as it was repeatedly pointed out, this
regime is dangerous to work in and as it has been done in \cite{svetitsky}
(see also \cite{billo} and \cite{aver}), we use the anisotropic lattice 
\begin{equation}
\beta _\tau \equiv \beta \equiv \frac{2N\xi }{g^2}>>\beta _\sigma \equiv 
\frac{2N}{g^2\xi }  \label{k}
\end{equation}
where $\xi =\sqrt{\frac{\beta _\tau }{\beta _\sigma }}$\ and $g^2=g_\tau
g_\sigma .$\ In the weak coupling region $g_\tau ^2\approx g_\sigma
^2\approx g^2+O\left( g^4\right) $ \cite{hasen1}, \cite{shig} so $\xi
\approx \frac{a_\sigma }{a_\tau }.$

As it is well known, the action for\ $Z\left( N\right) $\ gluodynamics in
such an approximation may be written as 
\begin{equation}
-S\approx -S_E=\beta \mathrm{Re}\sum_{\tau =0}^{N_\tau -1}\sum_{\mathbf{x}%
,\tau ,n}z_0\left( \mathbf{x},\tau \right) z_n\left( \mathbf{x},\tau
+1\right) z_0\left( \mathbf{x}+n,\tau \right) ^{*}z_n\left( \mathbf{x},\tau
\right) ^{*}  \label{Z-glu}
\end{equation}
and having$\ $imposed Hamiltonian gauge condition 
\begin{equation}
z_0\left( \mathbf{x},\tau \right) =1-\delta _\tau ^0+\delta _\tau ^0\Omega
\left( \mathbf{x}\right) ;\quad \Omega \left( \mathbf{x}\right) =\prod_{\tau
=0}^{N_\tau -1}z_0\left( \mathbf{x},\tau \right) ;\quad \Omega \left( 
\mathbf{x}\right) \in Z\left( N\right) ,
\end{equation}
we get 
\begin{equation}
-S_E=\beta \mathrm{Re}\sum_{\mathbf{x},n}\left\{ \Omega \left( \mathbf{x}%
\right) z_n\left( \mathbf{x},0\right) \Omega \left( \mathbf{x}+n\right)
^{*}z_n\left( \mathbf{x},N_\tau -1\right) ^{*}\right\} -S_{ch}  \label{z-act}
\end{equation}

The action $\left( \ref{z-act}\right) $\ is a set of one-dimensional chains 
\begin{equation}
-S_{ch}=\beta \mathrm{Re}\sum_{\mathbf{x},n}\left\{ \sum_{\tau =1}^{N_\tau
-1}z_n\left( \mathbf{x},\tau \right) z_n\left( \mathbf{x},\tau -1\right)
^{*}\right\}  \label{z-ch}
\end{equation}
\ attached by the first $z_n\left( \mathbf{x},0\right) $\ and last $%
z_n\left( \mathbf{x},N_\tau -1\right) $\ links to the plaquettes placed at $%
\tau =0.$ Beside such links, these plaquettes also contain Polyakov lines $%
\Omega \left( \mathbf{x}\right) $\ $.$\ It is evident that from 
\begin{equation}
e^{\beta \mathrm{Re}z}\equiv \sum_j\Im _j\left( \beta \right) z^j;\quad
\sum_{\left[ z\right] }z^j=\sum_{k=0}^{N-1}\exp \left\{ ij\varphi _k\right\}
=N\delta _0^j;\quad \varphi _k\equiv \frac{2\pi k}N
\end{equation}
follows\footnote{%
In the case of, e.g., $Z\left( 3\right) $ it gives: $\Im _0\left( \varkappa
\right) =\frac{e^\varkappa +2e^{-\frac \varkappa 2}}3$ and $\Im _{\pm
1}\left( \varkappa \right) =\frac{e^\varkappa -e^{-\frac \varkappa 2}}3$} 
\begin{equation}
\Im _j\left( \beta \right) =\frac 1N\sum_{\left[ z\right] }z^j\exp \left\{
\beta \mathrm{Re}z\right\} =\frac 1N\sum_{k=0}^{N-1}\exp \left\{ \beta \cos
\varphi _k+i\varphi _kj\right\} ,  \label{den}
\end{equation}
and having for each link 
\begin{eqnarray}
&&\exp \left\{ \tilde \beta \left( 1\right) \mathrm{Re}\left[ z_n\left( 
\mathbf{x},\tau \right) z_n\left( \mathbf{x},\tau -1\right) ^{*}\right]
\right\}  \nonumber \\
&=&\sum_{j=0}^{N-1}\Im _j\left( \beta \right) \cdot z_n\left( \mathbf{x}%
,\tau \right) ^jz_n\left( \mathbf{x},\tau -1\right) ^{*j}  \label{Z-1}
\end{eqnarray}
the sum over all variables except $z_n\left( \mathbf{x},0\right) $\ and $%
z_n\left( \mathbf{x},N_\tau -1\right) $\ we get the expression for $\left(
N_\tau -1\right) $-section chain 
\begin{eqnarray}
&&\ \ \sum_{\left[ z_n\left( \mathbf{x},\tau \right) \right] }^{\prime }\exp
\left\{ \tilde \beta \left( N_\tau -1\right) \mathrm{Re}\left[ \left( N_\tau
-1\right) z_n\left( \mathbf{x},\tau \right) z_n\left( \mathbf{x},\tau
-1\right) ^{*}\right] \right\}  \nonumber \\
\ &=&\sum_{j=0}^1\left[ \Im _j\left( \beta \right) \right] ^{N_\tau
-1}z_n\left( \mathbf{x},\tau \right) ^jz_n\left( \mathbf{x},\tau -1\right)
^{*j},  \label{Z-ch}
\end{eqnarray}
that differs from one-section chain simply by the substitution $\Im _j\left(
\beta \right) \rightarrow \left[ \Im _j\left( \beta \right) \right] ^{N_\tau
-1}.$\ Summing over the remaining chain variables one easily gets the
familiar expression 
\begin{eqnarray}
\exp \left\{ -S_{eff}\right\} &=&\Im _0\left( \beta \right) ^{N_\tau
}\sum_j\left( \frac{\Im _j\left( \beta \right) }{\Im _0\left( \beta \right) }%
\right) ^{N_\tau }\Omega \left( \mathbf{x}\right) ^j\Omega \left( \mathbf{x}%
+n\right) ^{*j}  \nonumber \\
\ &\approx &\Im _0\left( \beta \right) ^{N_\tau }\exp \left\{ \left( \frac{%
\Im _1\left( \beta \right) }{\Im _0\left( \beta \right) }\right) ^{N_\tau
}\Omega \left( \mathbf{x}\right) \Omega \left( \mathbf{x}+n\right)
^{*}\right\}  \label{Z-part}
\end{eqnarray}
and since only small $j$\ survive for $N_\tau \rightarrow \infty $ Eq. $%
\left( \ref{Z-part}\right) $\ leads to 
\begin{equation}
-S_{eff}\approx \tilde \beta \left( N_\tau \right) \sum_{\mathbf{x},n}\Omega
\left( \mathbf{x}\right) \Omega \left( \mathbf{x}+n\right) ^{*}-N_\tau F_0,
\label{Z-act}
\end{equation}
where the effective coupling $\tilde \beta \left( N_\tau \right) $\ is
completely defined by the one-dimensional chain and related to the bare
coupling $\beta $\ by the expression 
\begin{equation}
\left( \frac{\Im _1\left( \beta \right) }{\Im _0\left( \beta \right) }%
\right) ^{N_\tau }\approx \frac{\Im _1\left( \tilde \beta \left( N_\tau
\right) \right) }{\Im _0\left( \tilde \beta \left( N_\tau \right) \right) }%
\approx \tilde \beta \left( N_\tau \right) ;\quad N_\tau \rightarrow \infty
\label{rel}
\end{equation}
and the 'free' term\footnote{%
Only this term survives if we break the chain at $t=t_0$ with open boundary
conditions $z_0\left( \mathbf{x},t_0\right) \neq z_0\left( \mathbf{x}%
,t_0+N_\tau \right) .$ It easy to see that the first $N_\tau -1$ terms of
'high temperature' expansion contribute \textit{only} in $F_0$.} is 
\begin{equation}
-F_0\equiv \ln \Im _0\left( \beta \right)  \label{vac}
\end{equation}

In the wide area $\left( \beta \lesssim 1\right) $ the functions\ $\Im
_j\left( \beta \right) $\ $\approx I_j\left( \beta \right) $ (inter alia $%
\Im _j\left( \beta \right) \approx \frac{\beta ^j}{j!}$\ for $\beta <<1$)\
and therefore 
\begin{equation}
\tilde \beta \left( N_\tau \right) \approx \beta ^{N_\tau }  \label{str-c}
\end{equation}

However, in a very important area of $\beta >>N$\ the$\ \Im _j\left( \beta
\right) $\ values significantly differ from $I_j\left( \beta \right) $\ and
thus we obtain 
\begin{eqnarray}
\frac{\Im _1\left( \beta \right) }{\Im _0\left( \beta \right) } &\approx &%
\frac{1+2e^{-\beta \left( 1-\cos \frac{2\pi }N\right) }\cos \left( \frac{%
2\pi }N\right) }{1+2e^{-\beta \left( 1-\cos \frac{2\pi }N\right) }} 
\nonumber \\
&\approx &\exp \left\{ -2\left( 1-\cos \left( \frac{2\pi }N\right) e^{-\beta
\left( 1-\cos \frac{2\pi }N\right) }\right) \right\}
\end{eqnarray}
and, accordingly, for $N_\tau >>1$%
\begin{equation}
\tilde \beta \left( N_\tau \right) \approx \left( \frac{\Im _1}{\Im _0}%
\right) ^{N_\tau }\approx \exp \left\{ -2N_\tau \left( 1-\cos \left( \frac{%
2\pi }N\right) \right) e^{-\beta \left( 1-\cos \frac{2\pi }N\right)
}\right\} .  \label{ef-N}
\end{equation}
In particular 
\begin{equation}
\tilde \beta \left( N_\tau \right) \approx \left\{ 
\begin{array}{cc}
\exp \left\{ -4N_\tau e^{-2\beta }\right\} & N=2 \\ 
\exp \left\{ -3N_\tau e^{-\beta \frac 32}\right\} & N=3
\end{array}
\right.  \label{ef-c}
\end{equation}
so we come to $3d$\ $Z(N)$-spin model with an effective coupling given by $%
\left( \ref{ef-N}\right) .$\ Such models undergo phase transition at the
point\footnote{%
The critical point for $\left( \ref{Z-act}\right) $\ is defined as $\tilde
\beta \left( N_\tau \right) =\tilde \beta _c\left( N\right) $\ where $\tilde
\beta _c\left( N\right) $\ are numeric constants. In particular $\tilde
\beta _c\left( 2\right) \approx 0.2288$\ is the critical coupling of $d3$\
Ising model and $\tilde \beta $ $_c\left( 3\right) \approx 0.5501$\ $d3$ is
for Potts model.} $\tilde \beta \left( N_\tau \right) =\tilde \beta _c$\ .

\ Therefore for fixed $N_\tau $\ for the critical value of the bare coupling 
$\beta $ we get 
\begin{equation}
\beta _c\approx \frac{const+\ln \left( \frac 1{1-\cos \left( \frac{2\pi }%
N\right) }\right) }{1-\cos \left( \frac{2\pi }N\right) }  \label{B-C}
\end{equation}
which is similar to the well-known phenomenological rule \cite{bh-cr} 
\begin{equation}
\beta _c\approx \frac{const}{1-\cos \left( \frac{2\pi }N\right) }
\label{B-C-e}
\end{equation}

For $N>>1$\ we cannot cut the series in $j$\ at $\left( \ref{den}\right) $,
but we may expand $\cos \left( \frac{2\pi }Nj\right) \approx 1-\frac
12\left( \frac{2\pi }Nj\right) ^2$%
\begin{eqnarray}
\frac{\Im _j\left( \beta \right) }{\Im _0\left( \beta \right) } &\approx &%
\frac{\sum_{k=0}^{N-1}\exp \left\{ -\frac \beta 2\left( \frac{2\pi }N\right)
^2k^2+i\left( \frac{2\pi }N\right) kj\right\} }{\sum_{n=0}^{N-1}\exp \left\{
-\frac \beta 2\left( \frac{2\pi }N\right) ^2n^2\right\} } \\
&=&\frac{\theta _3\left( \frac \beta 2\left( \frac{2\pi }N\right) ^2;\left( 
\frac{2\pi }N\right) j\right) }{\theta _3\left( \frac \beta 2\left( \frac{%
2\pi }N\right) ^2;0\right) }.
\end{eqnarray}

Taking into account that for the elliptic function \cite{bateman} 
\begin{eqnarray}
\theta _3\left( \lambda ;\varphi \right)  &\equiv &\sum_{n=-\infty }^\infty
\exp \left\{ -\lambda n^2+in\varphi \right\}   \label{Th3} \\
\  &=&\sqrt{\frac \pi \lambda }\sum_{m=-\infty }^\infty \exp \left\{ -\frac{%
\left( \varphi +2\pi m\right) ^2}{4\lambda }\right\} ,  \nonumber
\end{eqnarray}
one can write with good accuracy \cite{BPZ} 
\begin{equation}
\theta _3\left( \lambda ;\varphi \right) \approx \exp \left\{ \mathtt{\tilde
\gamma }\left( \lambda \right) \cos \varphi +G\left( \lambda \right)
\right\} ,  \label{Th}
\end{equation}
where $\mathtt{\tilde \gamma }\left( \lambda \right) $\ and $G\left( \lambda
\right) $\ are smooth analytical functions of $\lambda $%
\begin{eqnarray}
\mathtt{\tilde \gamma }\left( \lambda \right)  &\approx &\left\{ 
\begin{array}{cc}
2\exp \left\{ -2\lambda \right\} ; & \lambda >>1 \\ 
\frac 1{2\lambda }; & \lambda <<1
\end{array}
\right. ;  \nonumber  \label{J-th} \\
G\left( \lambda \right)  &\approx &\left\{ 
\begin{array}{cc}
0 & \lambda >>1 \\ 
\frac 12\ln \frac \pi \lambda ; & \lambda <<1
\end{array}
\right.   \label{G-th}
\end{eqnarray}
and we get for $\tilde \beta \left( N_\tau \right) $%
\begin{equation}
-\ln \tilde \beta \left( N_\tau \right) \approx -\frac{N_\tau }{j^2}\ln 
\frac{\Im _j\left( \beta \right) }{\Im _0\left( \beta \right) }\approx \frac{%
N_\tau }2\left( \frac{2\pi }N\right) ^2\mathtt{\tilde \gamma }\left[ \frac
\beta 2\left( \frac{2\pi }N\right) ^2\right] ,
\end{equation}
which finally gives 
\begin{equation}
-\ln \tilde \beta \left( N_\tau \right) \approx 
\begin{array}{cc}
N_\tau \left( \frac{2\pi }N\right) ^2\exp \left\{ -\frac \beta 2\left( \frac{%
2\pi }N\right) ^2\right\} ; & \beta >>N
\end{array}
\label{fin}
\end{equation}
and 
\begin{equation}
-\ln \tilde \beta \left( N_\tau \right) \approx 
\begin{array}{cc}
\frac{N_\tau }{2\beta }; & \beta <<N
\end{array}
.  \label{inf}
\end{equation}

If we claim that the critical point $\tilde \beta \left( N_\tau \right)
=\tilde \beta _c$\ corresponds to the phase transition, the dependence of $%
\tilde \beta \left( N_\tau \right) $\ on $N_\tau $\ must be removed at least
in the continuum limit by a renormalisation procedure which relates $N_\tau $%
\ (or $a_\tau $\ ) to the bare coupling $\beta $. The conditions that
provide for such an independence are discussed later for some simple cases.

In the previous expressions we could not avoid somewhat annoying
'generalities', but we cherish hope that in what follows those may be
regarded as 'justified'.\ Finally we want to stress the fact that the
suggested approach is not limited to small $N_\tau $, however, the coupling $%
\tilde \beta \left( N_\tau \right) $\ in an effective action decrease with $%
N_\tau $\ so we must accordingly increase $\xi $\ to guarantee that $\tilde
\beta \left( N_\tau \right) >>\beta _\sigma $\ .

\section{Effective action for finite\ temperature $U\left( 1\right) $, $%
SU\left( 2\right) $ and $SU\left( 3\right) $ gluodynamics}

Wilson action for $SU(N)$ gauge group may be written in the static gauge $%
U_0\left( \mathbf{x},\tau \right) =U_0\left( \mathbf{x}\right) $ as 
\begin{equation}
-S_E=\frac \beta N\sum_{\tau =0}^{N_\tau -1}\sum_{\mathbf{x},,n}\mathrm{ReSp}%
\left\{ U_0\left( \mathbf{x}\right) U_n\left( \mathbf{x},\tau +1\right)
U_0^{\dagger }\left( \mathbf{x}+n\right) U_n^{\dagger }\left( \mathbf{x}%
,\tau \right) \right\}  \label{glu}
\end{equation}
and after the gauge transformation $U_n\left( \mathbf{x},\tau \right)
\rightarrow \left[ U_0\left( \mathbf{x}\right) \right] ^{-\tau }U_n\left( 
\mathbf{x},\tau \right) \left[ U_0\left( \mathbf{x}+n\right) \right] ^\tau $
we are sure to get \footnote{%
The same result may be obtained by imposing the Hamiltonian gauge condition $%
U_0\left( \mathbf{x},\tau \right) =1-\delta _\tau ^0+\delta _\tau ^0\Omega
\left( \mathbf{x}\right) $ with $\Omega \left( \mathbf{x}\right) \equiv
U_0\left( \mathbf{x}\right) ^{N_\tau }$, from the begining.} 
\begin{equation}
-S_E=\sum_{\mathbf{x},n}\left( \frac \beta N\mathrm{ReSp}\left\{ \Omega
\left( \mathbf{x}\right) U_n\left( \mathbf{x},0\right) \Omega ^{\dagger
}\left( \mathbf{x}+n\right) U_n^{\dagger }\left( \mathbf{x},N_\tau -1\right)
\right\} -s_{ch}\left( \mathbf{x},\mathbf{n}\right) \right)  \label{act}
\end{equation}
which, as well as $\left( \ref{Z-act}\right) ,$ is given by plaquettes,
placed on the last links in the temporal direction and represents $\Omega
\left( \mathbf{x}\right) $-loop interactions with the one-dimensional chains 
\begin{equation}
-s_{ch}\left( \mathbf{x},\mathbf{n}\right) =\frac \beta N\mathrm{ReSp}%
\left\{ \sum_{\tau =1}^{N_\tau -1}U_n\left( \mathbf{x},\tau \right)
U_n^{\dagger }\left( \mathbf{x},\tau -1\right) \right\}  \label{act-ch}
\end{equation}
attached to such plaquettes. Having written 
\begin{equation}
\exp \left\{ -s_{ch}\left( \mathbf{x},\mathbf{n}\right) \right\}
=\sum_jd_j\Im _j\left( \beta \right) \chi _j\left( U_n\left( \mathbf{x},\tau
\right) U_n^{\dagger }\left( \mathbf{x},\tau -1\right) \right) ,  \label{ch}
\end{equation}
where 
\begin{equation}
\Im _j\left( \beta \right) =\int \exp \left\{ \frac \beta N\mathrm{ReSp}%
\left\{ U_n\right\} \right\} \chi _j\left( U_n\right) dU_n  \label{coeff-U}
\end{equation}
and $\chi _j\left( U_n\right) $ is the characters of $j$ irreducible
representations. So using the character orthogonality 
\begin{eqnarray}
&&\ \ \int \chi _j\left( U_n\left( \mathbf{x},\tau \right) U_n\left( \mathbf{%
x},\tau ^{\prime \prime }\right) ^{\dagger }\right) \chi _k\left( U_n\left( 
\mathbf{x},\tau ^{\prime \prime }\right) U_n\left( \mathbf{x},\tau ^{\prime
}\right) ^{\dagger }\right) dU_n\left( \mathbf{x},\tau ^{\prime \prime
}\right)  \nonumber \\
\ &=&\frac{\delta _{jk}}{d_j}\chi _j\left( U_n\left( \mathbf{x},\tau \right)
U_n\left( \mathbf{x},\tau ^{\prime }\right) ^{\dagger }\right)  \label{ort}
\end{eqnarray}
and integrating over all $U_n\left( \mathbf{x},\tau \right) $ variables
except $U_n\left( \mathbf{x},0\right) $ and $U_n\left( \mathbf{x},N_\tau
-1\right) $ one can obtain for $\left( N_\tau -1\right) $-section chains 
\begin{eqnarray}
&&\ \ \int \exp \left\{ \frac \beta N\sum_{\tau =1}^{N_\tau -1}\mathrm{ReSp}%
\left\{ U_n\left( \mathbf{x},\tau \right) U_n^{\dagger }\left( \mathbf{x}%
,\tau -1\right) \right\} \right\} dU  \nonumber \\
\ &=&\sum_jd_j\Im _j\left( \beta \right) ^{N_\tau -1}\chi _j\left( U_n\left( 
\mathbf{x},0\right) U_n\left( \mathbf{x},N_\tau -1\right) ^{\dagger }\right)
,  \label{SU-ch}
\end{eqnarray}
which, as well as $\left( \ref{Z-ch}\right) ,$differs from a one-section
chain simply by $\Im _j\left( \beta \right) \rightarrow \Im _j\left( \beta
\right) ^{N_\tau -1}.$

Substituting$\left( \ref{SU-ch}\right) $ into $\left( \ref{act}\right) $ and
taking into account that 
\begin{equation}
\int \chi _j\left\{ \Omega \left( \mathbf{x}\right) U_n\left( \mathbf{x}%
,0\right) \Omega ^{\dagger }\left( \mathbf{x}+n\right) U_n^{\dagger }\left( 
\mathbf{x},0\right) ^{\dagger }\right\} dU=\frac{\chi _j\left( \Omega \left( 
\mathbf{x}\right) \right) \chi _j\left( \Omega \left( \mathbf{x}\right)
^{\dagger }\right) }{d_j}  \label{factor}
\end{equation}
we can integrate over $U_n\left( \mathbf{x},N_\tau -1\right) $ and finally
get the familiar result \cite{green,ogilvie} 
\begin{equation}
\exp \left\{ -\bar S_E\right\} =\sum_j\exp \left\{ -N_\tau \left( \lambda
_G^{\left( j\right) }\left( \beta \right) +\ln \Im _0\left( \beta \right)
\right) \right\} \chi _j\left( \Omega \left( \mathbf{x}\right) \right) \chi
_j\left( \Omega \left( \mathbf{x}\right) ^{\dagger }\right)  \label{Aj}
\end{equation}
with\footnote{%
In $\left( 1+1\right) $-dimensional case we can integrate over $\Omega
\left( \mathbf{x}\right) $ and wright for the partition function $%
Z=\sum_j\Im _j\left( \beta \right) ^{N_\sigma N_\tau }.$ This result is
exact because in this case the action does not contain the magnetic part.} 
\begin{equation}
\lambda _G^{\left( j\right) }\left( \beta \right) =\lambda _G^{\left(
j\right) }\left( \frac{2N}{g^2}\right) \equiv -\ln \frac{\Im _j\left( \beta
\right) }{\Im _0\left( \beta \right) }.  \label{Lj}
\end{equation}

For the gauge groups $U(1)$ and $SU(2)$ we shall have $\Im _j\left( \beta
\right) =I_j\left( \beta \right) $ and$\ \Im _j\left( \beta \right)
=I_{2j+1}\left( \beta \right) $ correspondingly. Having made allowance for (%
\cite{bateman}7.13.2$\left( 8\right) $) one can write 
\begin{equation}
I_n\left( \beta \right) \approx \left\{ 
\begin{array}{cc}
\frac 1{\sqrt{2\pi \beta }}e^\beta \exp \left\{ -\lambda n^2\right\} ; & 
\beta \gtrsim 1; \\ 
\frac{\left( \frac \beta 2\right) ^n}{n!}\exp \left\{ \left( \frac{1-\frac 1n%
}{4n}-\frac 1{18}\right) \beta ^2\right\} ; & \beta \lesssim 1.
\end{array}
\right.  \label{bess2}
\end{equation}

Therefore in the \textit{strong coupling region }$\left( \beta \lesssim
1\right) $, we get the result which is very similar to that obtained for $%
Z(N)$%
\begin{equation}
\lambda _G^{\left( j\right) }\left( \beta \right) =-\ln \frac{\Im _j\left(
\beta \right) }{\Im _0\left( \beta \right) }\backsim d_j\ln \left( \frac{2N}%
\beta \right) .  \label{coup U}
\end{equation}

Moreover, taking into account the results \cite{drouffe} obtained for $\int
\chi _j^{*}\left( U\right) e^{\beta \chi }dU$, Eq. $\left( \ref{coup U}%
\right) $ can be easily generalized for $SU(N)$ case. Therefore, we may
conclude that in the region $\beta \lesssim 1$ the effective coupling $%
\tilde \beta \left( N_\tau \right) $ depends on $\beta $ and $N_\tau $ as 
\begin{equation}
\tilde \beta \left( N_\tau \right) =F\left( \left( \frac \beta
{const}\right) ^{N_\tau }\right) .  \label{SUcombS}
\end{equation}

For the \textit{weak coupling region} we can write 
\begin{equation}
\lambda _G^{\left( j\right) }\left( \beta \right) \approx \lambda _Nd_A\cdot
\left[ C_2\left( j\right) +O\left( \frac j\beta \right) ^3\right] ,
\label{Cas}
\end{equation}
where $C_2\left( j\right) $is the quadratic Casimir operator, $d_A=N^2-1$
for $SU(N)$ and $d_A=1$ for $U(1)$.$\ $The function $\lambda _G$\ for $U(1)$%
\ is given by 
\begin{equation}
\lambda _1=-\ln \left( \frac{I_1\left( \beta \right) }{I_0\left( \beta
\right) }\right) \simeq \frac 1{2\beta }+\left( \frac 1{2\beta }\right)
^2\simeq \frac{g^2}4+\left( \frac{g^2}4\right) ^2  \label{coup1}
\end{equation}
\ and that for $SU(2)$ by 
\begin{equation}
\lambda _2=\frac 1{2\beta }+\left( \frac 1{2\beta }\right) ^2+O\left( \beta
^{-3}\right) =\frac{g^2}8+\left( \frac{g^2}8\right) ^2+O\left( g^6\right) .
\label{coup2}
\end{equation}

Similar, but less accurate result can be obtained for $SU(3)$ gauge group
(with $j=\left\{ l_1;l_2\right\} $). If we apply the asymptotic formula $%
\left( \ref{bess2}\right) $ to the well known expression \cite{drouffe} 
\begin{equation}
\int \exp \left\{ -\beta \frac \chi 3\right\} \chi _{l_1l_2}d\mu =\sum_n\det
I_{l_j-j+i-n}\left( \frac \beta 3\right)
\end{equation}
this will give 
\begin{eqnarray}
&&\int \exp \left\{ -\frac \beta 3\mathrm{ReSp}\left( U\right) \right\} \chi
_{l_1l_2}d\mu  \nonumber \\
&\approx &\det I_{l_j-j+i}\left( \frac \beta 3\right) e^{-\frac{8\lambda _3}%
3\left( l_1+l_2\right) ^2}\sum_n\exp \left\{ -8\lambda _3\left( n+\frac{%
l_1+l_2}{3\cdot }\right) ^2\right\} ,
\end{eqnarray}
with $l_2>l_1>0$ and 
\begin{equation}
\frac 83\lambda _3=\frac 1{\frac 23\beta }\left( 1+\frac 1{\frac 23\beta
}\right) +O(\beta ^{-3}).  \label{coup3}
\end{equation}
So, taking into account 
\begin{equation}
\sum_n\exp \left\{ -8\lambda _3\left( n-\frac{l_1+l_2}3\right) ^2\right\}
=\theta _3\left( 8\lambda _3;0\right) +O\left( \exp \left\{ -\frac{2\pi ^2}{%
38\lambda _3}\right\} \right) ,
\end{equation}
one can easily find 
\begin{eqnarray}
&&\int \exp \left\{ -\frac \beta 3\mathrm{ReSp}\left( U\right) \right\} \chi
_{l_1l_2}d\mu  \nonumber \\
&\approx &\sum_n\det I_{l_j-j+i-n}\left( \frac \beta 3\right) \approx \exp
\left\{ -8\lambda _3\cdot C_2\left( l_1,l_2\right) \right\} ,
\end{eqnarray}
where $C_2\left( l_1;l_2\right) =\frac 13\left( l_1^2+l_2^2-l_1l_2\right)
-\frac 13$ is the quadratic Casimir operator for $SU(3)$ group and we
immediately get an expression $\lambda _{SU(3)}^{\left( j\right) }\left(
\beta \right) $ that formally coincides with $\left( \ref{Cas}\right) .$
Therefore, the effective action is given by 
\begin{equation}
\exp \left\{ -\bar S_E\right\} \approx \left[ \Im _0\left( \beta \right)
\right] ^{N_\tau }\sum_je^{-8\lambda _3N_\tau \cdot C_2\left( j\right) }\chi
_j\left( \Omega \left( \mathbf{x}\right) \right) \chi _j\left( \Omega \left( 
\mathbf{x}\right) ^{\dagger }\right) .  \label{act3}
\end{equation}

In the first order in $\frac 1{\beta ^2}$ the expression $\left( \ref{act3}%
\right) $ coincides with that obtained in Hamiltonian approach developed in 
\cite{polyakov,susskind}. It is worth to note that in \cite
{polyakov,susskind} the Hamiltonian gauge $U_0\left( \mathbf{x},\tau \right)
=1$ can be imposed on all $\tau $-links (in distinction from $\left( \ref
{act}\right) $) because there are no periodic border conditions. As a rule
such condition is demanded within finite temperature theory \cite{gross},\
so at this point the approach of \cite{polyakov,susskind} resembles zero
temperature theory. However, as it is known from field theory \cite{slavnov}%
, if we fix the gauge in a such way we would destroy the mechanism that
guarantees the Gauss condition $Q^b$ $=0$ ($Q^b$ $=\left( \delta
^{bc}\partial _n-f^{abc}A_n^a\right) E_n^c$ for QFT and $Q^b$ $%
=\sum_{n=-d}^dE_n^b\ $for LGT) and a projection operator should be
introduced to restore it.

In field theory, as well as in LGT \cite{susskind} the integration variables 
$\phi _0^b\left( \mathbf{x}\right) $\ in the projection operator correspond
to the parameters of static gauge transformations with the generators $Q^b$ 
\cite{slavnov} and finally all looks as if the static gauge condition $%
U_0\left( \mathbf{x},\tau \right) =\mathrm{U}_0\left( \mathbf{x}\right)
=\exp \left\{ i\phi _0^b\left( \mathbf{x}\right) \mathrm{T}^b\right\} $ had
been imposed from the beginning\ (as in $\left( \ref{glu}\right) $) instead
of $U_0\left( \mathbf{x},t\right) =1$. Therefore, the auxiliary variables $%
\prod_{\tau =0}^{N_\tau }U_0\left( \mathbf{x},\tau \right) =\mathrm{U}%
_0\left( \mathbf{x}\right) ^{N_\tau }=\exp \left\{ iN_\tau \phi _0^b\left( 
\mathbf{x}\right) \mathrm{T}^b\right\} $\ in the projection operator
formally correspond to Polyakov loops $\Omega \left( \mathbf{x}\right) $\ in 
$\left( \ref{act3}\right) $\ and the lattice length in the temporal
direction $N_\tau a_\tau $\ corresponds to the inverse temperature. So, at
least in this particular case, the border conditions appear not to play a
significant role.

Thus, we can use the results \cite{BPZ} obtained in Hamiltonian approach to
sum over irreducible representation $j$ in $\left( \ref{Aj}\right) $ which,
e.g. for $SU(2)$ group, gives 
\begin{eqnarray}
&&\ \ \sum_je^{-j\left( j+1\right) \lambda }\cdot \frac{\sin \left[ \frac{%
\left( 2j+1\right) \varphi }2\right] \sin \left[ \frac{\left( 2j+1\right)
\varphi ^{\prime }}2\right] }{\sin \frac \varphi 2\sin \frac{\varphi \prime }%
2}  \nonumber \\
\ &=&e^{-\frac 14N_\tau \lambda }\cdot \frac{\theta _3\left( \frac \lambda 2;%
\frac{\varphi -\varphi ^{\prime }}2\right) -\theta _3\left( \frac \lambda 2;%
\frac{\varphi +\varphi \ ^{\prime }}2\right) }{2\sin \frac \varphi 2\sin 
\frac{\varphi ^{\prime }}2},\quad
\end{eqnarray}
and with good accuracy one can write 
\begin{eqnarray}
&&\ \ \frac{\theta _3\left( \frac \lambda 2;\frac{\varphi -\varphi ^{\prime }%
}2\right) -\theta _3\left( \frac \lambda 2;\frac{\varphi -\varphi ^{\prime }}%
2\right) }{2\sin \frac{\varphi _x}2\sin \frac{\varphi _{x+\nu }}2}  \nonumber
\\
\ &\approx &\exp \left\{ \mathtt{\tilde \gamma }\cos \frac \varphi 2\cos 
\frac{\varphi ^{\prime }}2+\delta \sin \frac \varphi 2\sin \frac{\varphi
\prime }2+G\right\} ,  \label{T2}
\end{eqnarray}
where $\mathtt{\tilde \gamma }\left( \frac \lambda 2\right) ,G\left( \frac
\lambda 2\right) $and $\delta \left( \frac \lambda 2\right) $ are smooth
analytical functions of $\lambda $ which are very similar to given in $%
\left( \ref{G-th}\right) $, and $\delta \left( \frac \lambda 2\right)
,G\left( \frac \lambda 2\right) <<\mathtt{\tilde \gamma }\left( \frac
\lambda 2\right) .$ Then we can finally write 
\begin{equation}
\exp \left\{ -\bar S_E\right\} \approx \exp \left\{ \tilde \beta \left(
N_\tau \right) \chi _j\left( \Omega \left( \mathbf{x}\right) \right) \chi
_j\left( \Omega \left( \mathbf{x}\right) ^{\dagger }\right) +const\right\} ,
\label{actFIN}
\end{equation}
with 
\begin{equation}
\tilde \beta \left( N_\tau \right) \equiv \mathtt{\tilde \gamma }\left( 
\frac{2N_\tau }\beta \right) \approx \left\{ 
\begin{array}{ll}
2\exp \left\{ -\frac{2N_\tau }\beta \right\} ; & N_\tau >>\beta >>1; \\ 
\frac \beta {N_\tau }; & \beta >>N_\tau .
\end{array}
\right.  \label{SU(2)comb}
\end{equation}

In similar manner one may show that in the case of $SU(3)$ gauge group
effective coupling dependence on $\beta $\ and $N_\tau $\ resembles $\left( 
\ref{SU(2)comb}\right) $\ and can be written as 
\begin{equation}
\tilde \beta \left( N_\tau \right) \approx f\left( \lambda _GN_\tau \right)
\label{SU(3)comb'}
\end{equation}
with $\lambda _G$\ given by $\left( \ref{coup1}\right) ,\left( \ref{coup2}%
\right) $\ and $\left( \ref{coup3}\right) .$

We may conclude that in given approximation for continuous abelian $U\left(
1\right) $\ and nonabelian groups $SU(2)$\ and $SU(3)$\ the effective
couplings $\tilde \beta \left( N_\tau \right) $\ are also totally determined
by one-dimensional chains, however, contrary to $Z(N)$\ case $\left( \ref
{fin}\right) $\ $\tilde \beta \left( N_\tau \right) $\ is changing \
essentially slower with the bare coupling $\beta $.

In Appendices A and B we made an attempt to perturbatively estimate the
contribution of the magnetic part of the action. First order correction in
magnetic coupling $\beta _\sigma $ turns into zero for symmetry reasons. The
most important (from our viewpoint) part of the second order correction $%
\beta _\sigma ^2\Xi _2$ leads to simple renormalisation of the effective
coupling and at least in a limit $N_\tau \rightarrow \infty $ does not
change the results drastically.

More complicated (and possibly more accurate) expression for $\beta _\sigma
^2\Xi _2$ was suggested in \cite{billo}. As it was established in this paper
the additional term $\beta _\sigma ^2\Xi _2$\ makes the movement of $T_c$\
with increasing $N_\tau $\ slower \cite{billo}. Unfortunately, it is not so
easy to anticipate whether it can totally terminate the dependence of $T_c$\
on\ $N_\tau ,$ especially on extremely large lattices, because the
perturbative expansion in $\beta _\sigma $ is hardly applicable outside of
the narrow area $\beta _\sigma ^2N_\tau \sim \frac 1T\frac{a_\tau }{g_\sigma
^4a_\sigma ^2}<<1$.

Moreover, if following \cite{billo} we assume that second order corrections
in $\beta _\sigma $\ radically change the predictions obtained from zero
order, the convergence of series in $\beta _\sigma $\ should be reconsidered
and one must prove that the corrections of orders higher than the second one
are insignificant. \texttt{A}nyway perturbative series can hardly be
regarded as a reliable instrument for the computation of effective action in
the area $\beta _\sigma \gtrsim 1$ , $N_\tau \rightarrow \infty $ (see also
footnote 4)

As it has been already pointed out, $\beta _\sigma =$\ $\frac{2N}{g^2\xi }$\
can be made negligibly small in comparison with $\beta _\tau =\frac{2N}{%
g^2\xi }$\ by changing the anisotropy parameter $\xi .$\ Precise analysis of
modifications of $\beta _{\tau ,\sigma }$\ with changing $\xi $\ has been
made by F. Karsch in~\cite{karsch2}: 
\begin{equation}
\beta _\tau =\xi (\bar \varkappa -0.27192)+\frac 12;~~~\beta _\sigma =\frac{%
\bar \varkappa +0.39832}\xi ;~~~\bar \varkappa =\sqrt{\beta _\tau \beta
_\sigma }.  \label{K}
\end{equation}
However, this going into details\texttt{\ }cannot change the situation
drastically for large$~\xi $\ (Hamiltonian limit) and large $\beta $\
(continuum limit)$.$\ In this particular area one can hardly hope that the
second order corrections in $\beta _\sigma $\ will play such a significant%
\texttt{\ }role.

\section{'Electric toy' model}

Since the pioneering work of Svetitsky and Yaffe~\cite{svetitsky}, it is
understood that all the relevant properties of the phase structure can be
encoded in a suitable effective action for the order parameter, e.g. the
Polyakov loop. The simplest and most popular approach to such an action was
suggested in \cite{green,ogilvie}. We use the anisotropic lattice ($a_\tau
<< $ $a_\sigma $ ), that helps to avoid the strong coupling approximation
(one of the basic assumption in \cite{green,ogilvie}).Then we try to trace $%
a_\tau $-dependance of the effective coupling and compute Callan-Symanzik
beta function for $N_\tau \rightarrow \infty $.

As it is pointed out in \cite{svetitsky} at high temperatures $T_{SY}$ 
\footnote{%
In \cite{svetitsky} temperature $T_{SY}$ is defined as $T_{SY}\equiv \sqrt{%
\varkappa _\tau /\varkappa _\sigma }/N_\tau a_\tau $} $\beta _\sigma
\backsim 1/T_{SY}$ and $\beta _\tau \backsim T_{SY}$ (in our denotation $%
\beta _\sigma \backsim 1/\xi $ and $\beta _\tau \backsim \xi ).$
Consequently, the functional integral is highly peaked at the configurations
in which the spatial link variables are static up to gauge transformations
and the system is described by zero temperature $d$-dimensional gauge theory
with coupling $g^2T_{SY}.$ The dynamics of the spatial gauge fields is
regarded \cite{svetitsky} as being qualitatively the same at high and low
temperatures.

The naive interpretation of this suggestion might be as follows: the
magnetic part of the action $S_M$, proportional to $\beta _\sigma $ $%
\backsim 1/\xi $ do not play an essential role at high temperatures and the
effective action, obtained by integration over spatial degrees of freedom
should not be influenced by the magnetic part in the vicinity of the
Hamiltonian limit $\left( \xi \rightarrow \infty \right) $. However, there
are serious reasons to believe that the magnetic part of the action may be
of crucial importance for creating the confining forces \cite{mack,yaffe,T},
even at high temperatures \cite{sim}.

Our computation reveals no drastic changes resulting from lower corrections
in $\beta _\sigma $ (see Appendices A and B), however, one would hardly
suggest with certainty that the picture will be the same for the whole
series in $\beta _\sigma $. A simple example would prove that such
suggestion may be erroneous. Indeed, let us consider $d2$ - Ising model on
the anisotropic lattice. Free energy density may be written as 
\begin{equation}
F\left( \gamma _\sigma ;\gamma _\tau \right) =-\frac{\ln Z}{N_\sigma N_\tau }%
=\ln \frac{\left( 1-\gamma _\sigma \right) \left( 1-\gamma _\tau \right) }%
2+\int_0^{2\pi }\frac{d\varphi _\tau d\varphi _\sigma }{\left( 2\pi \right)
^2}\ln f  \label{cont}
\end{equation}
where 
\begin{equation}
f=\left( 1+\gamma _\sigma ^2\right) \left( 1+\gamma _\tau ^2\right) -2\gamma
_\tau \left( 1-\gamma _\sigma ^2\right) \cos \varphi _\tau -2\gamma _\sigma
\left( 1-\gamma _\tau ^2\right) \cos \varphi _\sigma .  \label{integrand}
\end{equation}
and 
\begin{equation}
\gamma _{\sigma ,\tau }=\tanh \beta _{\sigma ,\tau }.
\end{equation}

It is evident that for any arbitrary small $\beta _\sigma $\ there is a
critical value of $\beta _\tau $%
\begin{equation}
\beta _\tau ^c=-\frac 12\ln \tanh \beta _\sigma  \label{sg}
\end{equation}
\ at which the system undergoes a phase transition. Although perturbative
expansion in $\beta _\sigma $\ gives a reasonable value of $F,$\ the finite
series in $\beta _\sigma $\ do not reproduce the singularity of $F$ at $%
\beta _\tau =\beta _\tau ^c$. In such simple case, the reason of the above
can be easily traced. The main contribution into integrand at$\left( \ref
{cont}\right) $ gives the area of $\varphi _{\sigma ,\tau }\backsim 0$\ , so 
\begin{equation}
\ln f=\ln \left( 1+\gamma _\tau -2\gamma _\tau \cos \varphi _\tau \right)
+v\gamma _\sigma -\frac 12\gamma _\sigma ^2v^2+O\left( \gamma _\sigma
^3\right)
\end{equation}
with 
\begin{equation}
v=\frac{1+\gamma _\tau +2\gamma _\tau \cos \varphi _\tau -2\left( 1-\gamma
_\tau \right) \cos \varphi _\sigma }{1+\gamma _\tau -2\gamma _\tau \cos
\varphi _\tau }\backsim \frac{1+5\gamma _\tau }{1-\gamma _\tau }
\end{equation}
hence, in the area of $\gamma _\tau \sim 1$ the expansion parameter of free
energy $F$\ is, indeed, $\frac 6{1-\gamma _\tau }\gamma _\sigma $ rather
then $\gamma _\sigma $.

In the considered example, the restoration of real physical picture with
extending perturbative series in $\beta _\sigma $ is very slow. For this
reason, one should be careful when cutting the series in $\beta _\sigma $ in
gluodynamics; at some stage the magnetic part contribution may start
dominating and might completely change the final result. Therefore, $QCD$
without the magnetic part should be considered as a specific model.

An effective action $\left( \ref{Aj}\right) $ of such \textit{electric toy
model} differs from that obtained in \cite{green,ogilvie} only by initial
assumption: we do not use strong coupling approximation and work on the
anisotropic lattice. An additional assumption that the magnetic part may be
discarded\texttt{\ }in the limit $\xi \rightarrow \infty $ ,\texttt{\ }give
us the right to study weak coupling region and consider the limit $a_\tau
\rightarrow 0$\ in such model. It is evident, that at the same time we can
demand $a_\sigma \rightarrow 0$ and $a_\tau N_\tau =T^{-1}=const<\infty .$

Although $\left( \ref{Aj}\right) $ was computed exactly along the line of
Svetitsky-Yaffe program, discarded magnetic part of the action may lead to
serious distortions. Therefore, it is interesting to compare $\left( \ref{Aj}%
\right) $ with the analytical results obtained without such approximation.
In \cite{svetitsky} the analytical solution was found for $(2+1)$%
-dimensional $U(1)$ -gluodynamics and a relation was established between $%
(2+1)$-dimensional $U(1)$-gluodynamics and $2$-$d$ Coulomb Gas model (CGM),
which undergoes a Berezinskii-Kosterlitz-Thoulessf phase transition at the
point of $2\pi ^2\frac \beta {N_\tau }=\beta _c^{CGM}\approx \frac 2\pi $.

The model action $\left( \ref{Aj}\right) $\ connects $(2+1)$-dimensional $%
U(1)$\ -gluodynamics with $2d$-$XY$\ model. As it is known, the 2D $X$-$Y$\
model can be mapped to the 2D Coulomb gas (for recent review see \cite{pier}%
). Therefore, in exact \cite{svetitsky} and approximate approach one comes
to similar description of the phase structure $(2+1)$- dimensional $U(1)$%
-gluodynamics. The only thing which remains is to compare the prediction of 
\cite{svetitsky} for the critical coupling $\beta _c\approx \frac 4{\pi
T}\approx \frac{1.2732}T$ (in $a_\tau =1$units ) with that one which follows
for $N_\tau >>\beta $ from $\left( \ref{Aj}\right) $, that is with $\beta
_c\approx \frac{\beta _c^{BKT}}T$ (in the same units), where $\beta _c^{BKT}$
is the critical coupling of Berezinskii-Kosterlitz-Thoulessf phase
transition. Mote Carlo experiment \cite{BC-HMP} gives $\beta _c^{BKT}\approx
1.11.$ For the \textit{\ toy }model, we cannot expect more.

\section{Effective coupling and critical temperature}

The effective action $\left( \ref{Z-act}\right) $ for $Z\left( N\right) $ -
gluodynamics has the critical point at $\tilde \beta \left( N_\tau \right)
=\tilde \beta _c\left( N_\tau \right) $ where $\tilde \beta _c\left( N_\tau
\right) $ are numeric constants and the critical point will depend only on $%
N $ and the space dimension. If we assume that this critical point does
correspond to the temperature phase transition, then as it follows from $%
\left( \ref{str-c}\right) $ this will happen at the fixed temperature $T_c$
(in the strong coupling regime) when 
\begin{equation}
\left( \frac \beta {const}\right) ^{N_\tau }=const,
\end{equation}
then by the appropriate choice of constants one can obtain scaling having
imposed the condition $\left( \ref{g-gap}\right) $, but as for the string
tension the scaling is hardly possible under $\left( \ref{g-loop}\right) $
condition.

In the\textit{\ weak coupling} area, in accordance with $\left( \ref{fin}%
\right) ,$\ and $\left( \ref{ef-c}\right) $ we may put 
\begin{equation}
\tilde \beta \left( N_\tau \right) \approx \left\{ 
\begin{array}{cc}
\exp \left\{ -4N_\tau e^{-2\beta }\right\} ; & N=2; \\ 
\exp \left\{ -3N_\tau e^{-\beta \frac 32}\right\} ; & N=3; \\ 
\exp \left\{ -N_\tau \left( \frac{2\pi }N\right) ^2\exp \left\{ -\frac \beta
2\left( \frac{2\pi }N\right) ^2\right\} \right\} ; & N>>1.
\end{array}
\right.
\end{equation}
so as it follows from 
\begin{eqnarray}
-\ln \tilde \beta \left( N_\tau \right) &\approx &N_\tau \left( \frac{2\pi }%
N\right) ^2e^{-\frac \beta 2\left( \frac{2\pi }N\right) ^2}  \nonumber \\
&=&N_\tau \left( \frac{2\pi }N\right) ^2\left( \Lambda a_\tau \right) ^{\xi 
\frac{b_0\left( 4\pi ^2\right) }N}\approx -\ln \tilde \beta _c\approx const,
\end{eqnarray}
the critical point (e.g. for $N>>1$) will escape undesirable dependance on $%
N_\tau $\ if we demand 
\begin{equation}
\frac \beta {2N}\equiv g_\tau ^{-2}\approx b_0\ln \frac 1{\Lambda a_\tau
};\quad \Lambda =const  \label{Z-ren}
\end{equation}
where 
\begin{equation}
\ b_0\approx \frac N{4\pi ^2}\approx .025N\ ,  \label{b}
\end{equation}
which for $\xi =1$ and $N$ gives good agreement with $b_0=\frac{11}{12}\frac
N{4\pi ^2}\approx 0.023N$ . Therefore, we may conclude that 
\begin{equation}
T_c\simeq const\times a_\tau ^{-1+\xi \frac{b_0\left( 4\pi ^2\right) }%
N}\simeq const\times a_\tau ^{-1+\xi }.  \label{tem}
\end{equation}

As it has been shown \cite{petrov} 
\begin{equation}
\sqrt{\sigma }\approx \frac{a^{-1}}{\sqrt{2}}\exp \left\{ -\beta \left(
1-\cos \frac{2\pi }N\right) \right\} \approx \frac{a^{-1}}{\sqrt{2}}\exp
\left\{ -\frac{4\pi ^2}Nb_0\ln \frac 1{\Lambda a_\tau }\right\}
\label{sigma}
\end{equation}
and $\left( \ref{tem}\right) $ at given approximation the scaling will take
place or, in other words, we shall have 
\begin{equation}
\frac{\sqrt{\sigma }}{T_c}=\left( a_\tau \right) ^{\frac{b_0\left( 4\pi
^2\right) }N\left( \xi -\frac 1\xi \right) }\cdot const.  \label{sc}
\end{equation}
In principle, by changing $\xi $ one can tune the parameter $b_0$ to make $%
T_c$ in $\left( \ref{tem}\right) $ or $\sqrt{\sigma }$ in $\left( \ref{sigma}%
\right) $ tend to the finite value in the continuum limit, however, at
least, under given approximation it cannot be done simultaneously and the
ratio $\frac{\sqrt{\sigma }}{T_c}$ will be independent from lattice spacing
only for $\xi =1$.

\section{Continuous groups}

In this case the requirement that the position of critical point $\tilde
\beta \left( N_\tau \right) =\tilde \beta _c$ corresponding to the
temperature $T=T_c$ of phase transition should be independent of lattice
sizes, (at least for $N_\tau \rightarrow \infty )$ leads to the condition 
\begin{equation}
N_\tau \lambda _G\rightarrow const.  \label{cond}
\end{equation}
For large enough $\beta $\ (small $a_\tau $) $\lambda _{zG}\sim \frac
1{2\beta },$\ so $\left( \ref{cond}\right) $\ means trivial asymptotic
freedom 
\begin{equation}
\frac 1{2N}\lim_{a_\tau \rightarrow 0}\frac{g^2}{a_\tau }\equiv T_N=const.
\label{triv}
\end{equation}
For example, in the case of $SU(2)$ gauge group from $\left( \ref{cond}%
\right) $ and $\left( \ref{SU(2)comb}\right) $ we obtain the condition 
\begin{equation}
\tilde \beta \left( N_\tau \right) \approx \mathtt{\tilde \gamma }\left( 
\frac{N_\tau }Ng^2\right) \approx \mathtt{\tilde \gamma }\left( \frac 1N%
\frac{T_N}{T_c}\right) \approx \tilde \beta _c
\end{equation}
and consequently 
\begin{equation}
T_N=T_cN\mathtt{\tilde \gamma }^{-1}\left( \tilde \beta _c\right) ,
\end{equation}
where $\mathtt{\tilde \gamma }^{-1}\left( x\right) =y$ is the function
inverse to $\mathtt{\tilde \gamma }\left( y\right) =x,$ so $\mathtt{\tilde
\gamma }^{-1}\left( \tilde \beta _c\right) $ is simply a numeric constant.

Now we may write for the effective coupling 
\begin{equation}
\tilde \beta \left( N_\tau \right) \approx \mathtt{\tilde \gamma }\left( 
\mathtt{\tilde \gamma }^{-1}\left( \tilde \beta _c\right) \frac{T_c}T\right)
\end{equation}
with 
\begin{equation}
\mathtt{\tilde \gamma }\left( \mathtt{\tilde \gamma }^{-1}\left( \tilde
\beta _c\right) t\right) \approx \left\{ 
\begin{array}{ll}
2\exp \left\{ -t\mathtt{\tilde \gamma }^{-1}\left( \tilde \beta _c\right)
\right\} ; & t>>1; \\ 
\frac t{2\mathtt{\tilde \gamma }^{-1}\left( \tilde \beta _c\right) }; & t<<1.
\end{array}
\right.
\end{equation}

For the calculation of temporal string tension $\alpha $, higher orders in $%
a_\tau $\ are needed in the expansion of $g^2.$\textrm{\ }Indeed, to find $%
\alpha $ we may compute the correlation function $\left\langle \chi _0\chi
_R\right\rangle $ between two probes\footnote{%
Distance $R$ is measured in lattice units.} 
\begin{equation}
\left\langle \chi _0\chi _R\right\rangle =\int \frac{\exp \left\{
iqR\right\} }{s_0-\gamma \sum_n\cos q_n}\left( \frac{dq}{2\pi }\right)
^3;\quad \gamma =\frac{g^2}{2aT}.
\end{equation}
That can be done in spherical model approximation in the same way as in \cite
{BPZ}. The saddle point $s_0$ is defined from the condition $\left\langle
\chi _0\chi _R\right\rangle =1$ and is equal to 
\begin{equation}
s_0\approx \left\{ 
\begin{array}{cc}
3\gamma _c+8\pi ^2\gamma _c\left( \gamma _c-\gamma \right) ^2; & \gamma
\lesssim \gamma _c.
\end{array}
\right.
\end{equation}
For $R>>1$ one may write 
\begin{equation}
\left\langle \chi _0\chi _R\right\rangle \approx \left( 2\pi R\right)
^{-\frac 32}\exp \left\{ -\alpha R\right\}
\end{equation}
and for string tension $\alpha $ we get 
\begin{equation}
\alpha \sim \left( \gamma _c-\gamma \right) \cdot \theta \left( \gamma
_c-\gamma \right)
\end{equation}
or 
\begin{equation}
\alpha \approx \frac{g^2\left( T\right) -g^2\left( T_c\right) }{2a^2T}\theta
\left( T_c-T\right) .  \label{str-T}
\end{equation}
If we claim the independence of $\alpha $ from lattice sizes then we should
demand: 
\begin{equation}
g^2\left( T\right) \approx \left\{ 
\begin{array}{cc}
-4Ta\ln \left( \exp \left\{ -\frac{g^2\left( T_c\right) }{4Ta}\right\}
-\frac \alpha 2a\right) ; & T<T_c.
\end{array}
\right.  \label{g-T}
\end{equation}

This expression guarantees scaling for $\alpha $ in given approximation.
However, if in continuum limit $\alpha \rightarrow const$ then, at least,
the second order expansion $g^2T$ in $a$ is needed and, consequently, in
this case we must compute $\beta _{CS}\left( g^2\right) $ in the higher
orders in $g$. An example of a such computation is given in the Appendix C.

As it is seen from Eq. $\left( \ref{beta2}\right) ,$ $\beta _{CS}\left(
g^2\right) \sim -\frac g2$ when $g\rightarrow 0$, which strongly contradict
to standard expression $\left( \ref{bet}\right) ,$where $\beta _{CS}\left(
g^2\right) \sim -b_0g^3$ in corresponding area. It looks so as if for
extremely small $a$ (and consequently extremely small $g^2\left( a\right) $
) the 'recursive' coupling $g^2\left( \lambda a\right) \equiv F\left(
g^2(a);\lambda \right) $ might be presented as 
\begin{equation}
g^2(\lambda a)\simeq c_0\left( \lambda \right) +c\left( \lambda \right)
\cdot (g^2(a))^\alpha +...
\end{equation}
where $\alpha $ is not obligatory an integer, but is \textit{independent} of 
$\lambda $. Asymptotic freedom leads to $c_0\left( \lambda \right) =0$, so
we immediately get 
\begin{equation}
\beta _{CS}\left( g\right) =a\frac{\partial g}{\partial a}\simeq \frac{%
c\left( 1\right) }{c^{\prime }\left( 1\right) }g+O(g^3).
\end{equation}
Although the phase structure of the system defined by the action $\left( \ref
{glu}\right) $ and $\left( \ref{Aj}\right) $ essentially depends upon space
dimension $d,$ the effective coupling is not sensitive to it and is defined
by one-dimensional chain. Therefore, it is not too surprising that the toy
model shows \textit{trivial} asymptotic freedom inherent to Schwinger model%
\footnote{%
Another cause of similarity is the absence of magnetic field in Schwinger
model.}.

\section{Conclusions}

Although in a strong coupling region non-universal behavior of $\beta
_{CS}\left( g\right) $ is observed, in a weak coupling limit we find the
dependence of $g^2$ on lattice spacing for $Z(2)$ and $Z(3)$ gauge groups.
This dependence is very close to that obtained in standard renormalisation
theory. However, the dependence on lattice anisotropy parameter $\xi $ is
too strong compared with \cite{karsch2} and differs for spatial string
tension and critical temperature. Therefore, either an additional procedure
must be added to remove the remaining $\xi $-dependence in physical values
or one must work only at $\xi =1.$ In the last case our results should be
reconsidered, because we worked far from the area of $\xi \simeq 1.$

Main approximation, used to obtain an effective action for continuous groups
is that in a Hamiltonian limit $\left( \xi >>1\right) $ the magnetic part of
the action, (proportional to $1/\xi $) can be neglected compared to the
electric one (proportional to $\xi $). Perturbative estimation of the
magnetic part does not change the results considerably.\ Therefore,in the
limit of $N_\tau \rightarrow \infty $ \ and $a_\tau \rightarrow 0$ the toy
model shows the trivial asymptotic freedom $\left( g^2\sim a_\tau \right) .$
This strongly contradict the results obtained in the perturbation theory.

Another nameworthy point is to be underscored: a considerable difference
between $Z(N)$\ and continuous groups claimed by equations $\left( \ref
{Z-ren}\right) $ and $\left( \ref{beta2}\right) .$ It looks somewhat
ridiculous so we would like to discuss it in more detail. We argue that such
difference between $Z(N)$\ and $U(1)$ will not disappear\ for any large (but
finite) $N$\ and will be washed out only at $N\rightarrow \infty $\ ,
because on their way to continuum limit $\left( \beta \rightarrow \infty
\right) $\ $Z(N)$\ groups inevitably pass the point of $\beta \backsim \frac{%
N^2}{\left( 2\pi \right) ^2}=const.$ With further decreasing $a$ the bare
coupling $\beta >>\frac{N^2}{\left( 2\pi \right) ^2}$\ and, consequently,
effective coupling $\tilde \beta \left( N_\tau \right) $ exponentially
decrease with $\beta ,$\ as it can be seen from $\left( \ref{ef-c}\right) $
which \ finally leads to $\left( \ref{beta2}\right) ,$ so the dependence of $%
g^2$ on $a$ is similar to standard and leads to nontrivial asymptotic
freedom. In the case of $U(1)$ gauge group (which\ corresponds to $N=\infty $%
) the point of $\beta \backsim \frac{N^2}{\left( 2\pi \right) ^2}$ is
evidently unreachable. Therefore, we get different results if the limit $%
N\rightarrow \infty $ is taken before the computation of partition function,
and after that, in other words, the limits $N\rightarrow \infty $ and $%
N_\tau \rightarrow \infty $ do not 'commutate'.

Finally we want to stress again that despite the gluodynamics without a
magnetic part is very far from reality, it may be interesting not only for
simplicity reasons, but also as an example of a model where the
renormalisation procedure may be fulfilled analytically and exclusively
within the lattice gauge theory. At the same time, it is a good laboratory
to study the nature of some non-perturbative phenomena in LGT revealing the
difference between the models with discrete and continuous gauge groups and
it is not unlikely that this difference will be preserved in more realistic
models.

\vspace{1.0in}%
%
%
%
%

\section{Appendix A. Perturbation series in $\beta _\sigma \equiv \frac{2N}{%
g^2\xi }$ for $Z(2)$\ gauge group.}

Now, for simplicity, we shall confine ourself to the case of $Z(2)$\ gauge
group and, in fact, use the series in $\mathtt{\ }\gamma _\sigma \equiv
\tanh \beta _\sigma $: 
\begin{equation}
Z\left( \Omega \right) =Z_0\left( \Omega \right) \cdot \sum_{n=0}^\infty
\gamma _\sigma ^n\Xi _n\left( \Omega \right)
\end{equation}
The zero order term $Z_0\left( \Omega \right) $has been already considered.
It is easy to see that the first order term $\Xi _1$ is equal to zero. Only
the pairs of plaquettes having equal spatial coordinates and orientations,
but positioned at different temporal points $\ \tau $ and $\ \tau +\Delta $
along the temporal axis contribute into the second order $\Xi _2$. Each of
the four chains $1+\Omega \left( \mathbf{x}\right) \Omega \left( \mathbf{x+n}%
\right) \gamma ^{N_\tau }$ which cross pair links $z_n\left( \tau ,\mathbf{x}%
\right) $ and $z_n\left( \tau +\Delta ,\mathbf{x}\right) $ of these
plaquettes is converted into $\gamma ^\Delta +\Omega \left( \mathbf{x}%
\right) \Omega \left( \mathbf{x+n}\right) \gamma ^{N_\tau -\Delta }$ .

If we denote 
\begin{eqnarray}
\Omega \left( \mathbf{x}\right)  &=&\Omega _1;\quad \Omega \left( \mathbf{x+n%
}\right) =\Omega _2;  \nonumber \\
\Omega \left( \mathbf{x+n+m}\right)  &=&\Omega _3;\quad \Omega \left( 
\mathbf{x+m}\right) =\Omega _4
\end{eqnarray}
and 
\begin{eqnarray}
\mathbf{\hat I} &\equiv &\frac{\Omega _1\Omega _2+\Omega _2\Omega _3+\Omega
_3\Omega _4+\Omega _4\Omega _1}4;  \label{J} \\
\mathbf{\hat I}^2 &=&\frac{1+\Omega _1\Omega _3+\Omega _2\Omega _4+\Omega
_1\Omega _2\Omega _3\Omega _4}4  \nonumber
\end{eqnarray}
and take into account that $\mathbf{\hat I}=\mathbf{\hat I}^3$ and 
\begin{equation}
\prod_{n=1}^4\left( p+q\Omega _n\Omega _{n+1}\right) =\left( p^2-q^2\right)
^2+4pq\left( p^2+q^2\right) \mathbf{\hat I}\frak{+}8p^2q^2\mathbf{\hat I}^2
\end{equation}
after bulky computations we get \cite{aver}: 
\begin{equation}
\Xi _2\left( \Omega \right) =N_\tau \sum_{\Delta =1}^{N_\tau -1}\prod_{n=1}^4%
\frac{\gamma ^\Delta +\Omega _n\Omega _{n+1}\gamma ^{N_\tau -\Delta }}{%
1+\Omega _n\Omega _{n+1}\gamma ^{N_\tau }}\equiv N_\tau ^2\left( Q_0+2Q_1%
\mathbf{\hat I}+Q_2\mathbf{\hat I}^2\right)   \label{sec}
\end{equation}
where $Q_j$\ are the functions of $\gamma $\ and $N_\tau $. If e.g. for $Z(2)
$\ we denote 
\begin{equation}
\gamma =\exp \left\{ -2e^{-2\beta }\right\} =\exp \left\{ -2\left( a_\tau
\Lambda \right) ^{8\tilde b_0}\right\} 
\end{equation}
then for $8\tilde b_0\approx 1$\ it gives $\gamma ^{N_\tau }\approx
e^{-\varepsilon }$\ with $\varepsilon =\frac{2\Lambda }T$\ and for $N_\tau
>>1$\ we obtain 
\begin{eqnarray}
Q_0 &\approx &\frac{\sinh 2\varepsilon -2\varepsilon }{4\varepsilon \sinh
^2\varepsilon }\simeq \left\{ 
\begin{array}{cc}
\frac 13-\frac 2{45}\varepsilon ^2; & \varepsilon <<1; \\ 
\frac 1{2\varepsilon }; & \varepsilon >>1;
\end{array}
\right. \quad  \\
Q_1 &\approx &\frac{\left( \cosh 2\varepsilon +5\right) \sinh \varepsilon
-6\varepsilon \cosh \varepsilon }{4\varepsilon \sinh ^4\varepsilon }\simeq
\left\{ 
\begin{array}{cc}
\frac 15-\frac{17\varepsilon ^2}{210}; & \varepsilon <<1; \\ 
\frac 1\varepsilon e^{-\varepsilon }; & \varepsilon >>1;
\end{array}
\right.  \\
Q_2 &\approx &\frac{\left( 2\varepsilon \cosh \varepsilon -3\sinh
\varepsilon \right) \cosh \varepsilon +\varepsilon }{8\varepsilon \left(
\sinh \varepsilon \right) ^4}\simeq \left\{ 
\begin{array}{cc}
\frac 1{30}-\frac{\varepsilon ^2}{63}; & \varepsilon <<1; \\ 
\frac 14e^{-2\varepsilon }; & \varepsilon >>1.
\end{array}
\right. 
\end{eqnarray}

The term $\mathbf{\hat I}$ contains only the nearest neighbor interactions
of $\Omega _n$ and leads to a simple shift of effective coupling. Although
the term $\mathbf{\hat I}^2$ also includes interactions, that are absent in
the standard Ising model, such modifications has been extensively studied
and it was shown (see e.g.\cite{alonso} ) that they might change the phase
structure so drastically that this theory would not belong any more to the
same universality class as the ordinary one. However, if the corresponding
coupling $\tilde \beta \left( N_\tau \right) -\tilde \beta \left( N_\tau
\right) $is less than $\tilde \beta \left( N_\tau \right) /6$ (as it
certainly is in our case) such interactions does not bring sufficient
changes into the critical behavior of partition function \cite{alonso}.
Taking into account that only in the $1/8$ part of the configurations $%
\left( \Omega _n=-\Omega _{n+1}\right) ,$ $\mathbf{\hat I}$ differs from $%
\mathbf{\hat I}^2$ we put $Q_0+2Q_1\mathbf{\hat I}+Q_2\mathbf{\hat I}%
^2\approx Q_0+\left( 2Q_1+Q_2\right) \mathbf{\hat I}$ . It is easy to check
that in such approximation $K+\varepsilon \mathbf{\hat I}\approx K\exp
\left( \frac \varepsilon K\cdot \mathbf{\hat I}\right) $ so after the
inclusion of the second order correction we finally get 
\begin{eqnarray}
\frac{Z\left( \Omega \right) }{Z_0\left( \Omega \right) } &\approx &\left(
1+\gamma _\sigma ^2N_\tau ^2Q_0\right) \left( 1+\frac{\gamma _\sigma
^2N_\tau ^2\left( 2Q_1+Q_2\right) }{1+\gamma _\sigma ^2N_\tau ^2Q_0}\mathbf{%
\hat I}+...\right)  \nonumber \\
\ &\approx &\exp \left( \tilde \beta _{\left( 2\right) }\left( N_\tau
\right) \mathbf{\hat I}+const\right)
\end{eqnarray}
with 
\begin{equation}
\tilde \beta _{\left( 2\right) }\left( N_\tau \right) =\frac{2Q_1+Q_2}{%
\left( \gamma _\sigma N_\tau \right) ^{-2}+Q_0}  \label{shift2}
\end{equation}
which\ evidently leads to a simple shift in effective coupling.

\section{Appendix B. Perturbation series in $\beta _\sigma $ for $U\left(
1\right) $ - gauge group}

The zero order term $Z_0\left( \Omega \right) $ of the expansion 
\begin{equation}
Z\left( \Omega \right) =Z_0\left( \Omega \right) \cdot \left( 1+\beta
_\sigma \Xi _1\left( \Omega \right) +\frac{\beta _\sigma ^2}2\Xi _2\left(
\Omega \right) +...\right)
\end{equation}
corresponds to the case $\beta _\sigma =0$ already considered. It is easy to
see that the first order term $\Xi _1$ is equal to zero. Computation of the
second order term: $\Xi _2$ is very similar to that for discrete groups \cite
{aver}, but technical difficulties increase enormously. The procedure
suggested here differs from that elaborated in \cite{billo} chiefly in
technicalities. We are forced to partly sacrifice accuracy to obtain a
result which appears to us rather simple and transparent.

Only the pairs of plaquettes having equal spatial coordinates and
orientations contribute into the second order term: $\Xi _2$ but they are
positioned at different points $\tau $ and $\tau +\Delta $ along the
temporal axis.

Such additional plaquettes convert one-dimensional chains 
\begin{equation}
\sum_j\ \ \left[ I_j\left( \beta \right) \right] ^{N_\tau }e^{ij\left(
\varphi _{\mathbf{x}}-\mathbf{\varphi }_{\mathbf{x+n}}\right) },
\end{equation}
\ which cross the pair links $U_n\left( \tau ,\mathbf{x}\right) $\ and $%
U_n\left( \tau +\Delta ,\mathbf{x}\right) $\ of these plaquettes into 
\begin{equation}
\sum_j\left[ I_{j+1}\left( \beta \right) \right] ^\Delta \ \left[ I_j\left(
\beta \right) \right] ^{N_\tau -\Delta }e^{ij\left( \varphi _{\mathbf{x}}-%
\mathbf{\varphi }_{\mathbf{x+n}}\right) },
\end{equation}
so we get 
\begin{equation}
\Xi _2=\frac{N_\tau \sum_{jlrs}\Theta {}_{jrls}}{\sum_{jlrs}\ \ \left( \frac{%
I_j\left( \beta \right) I_r\left( \beta \right) I_l\left( \beta \right)
I_s\left( \beta \right) }{I_0\left( \beta \right) ^4}\right) ^{N_\tau }\cos
\left( \phi _{j,r,l,s}\right) }
\end{equation}
where 
\begin{equation}
\Theta {}_{jrls}=2\cos \left( \phi _{j,r,l,s}\right) \sum_{\Delta =0}^{\frac{%
N_\tau }2-1}\ \left( \frac{I_{j+1}I_{r+1}I_{l-1}I_{s-1}}{I_0^4}\right)
^\Delta \left( \frac{I_jI_rI_lI_s}{I_0^4}\right) ^{N_\tau -\Delta }
\end{equation}
and 
\begin{eqnarray}
\phi _{jrls} &=&j\left( \varphi _{\mathbf{x}}-\varphi _{\mathbf{x+n}}\right)
+r\left( \varphi _{\mathbf{x+n}}-\varphi _{\mathbf{x+n+m}}\right) + 
\nonumber \\
&&l\left( \varphi _{\mathbf{x+n+m}}-\varphi _{\mathbf{x+m}}\right) +s\left(
\varphi _{\mathbf{x+m}}-\varphi _{\mathbf{x}}\right) .
\end{eqnarray}

The term $\Xi _2$ may be easily computed for $\frac{N_\tau }{2\beta }\gtrsim
1$ (e.g., $N_\tau \sim \frac 1{a_\tau }$; $\beta \sim \ln \frac 1{a_\tau }$; 
$a_\tau \rightarrow 0$).\ In such area, taking into account $\left( \ref
{bess2}\right) ,$ leading terms may be written as 
\begin{equation}
\Theta _{0000}=\Theta _{-1-111}=2\frac{1-e^{-\frac{2N_\tau }\beta }}{%
1-e^{-\frac 4\beta }}
\end{equation}
The next terms 
\begin{eqnarray}
\Theta _{\left( 1\right) } &=&2\left( \Theta _{-1000}+\Theta _{0-100}+\Theta
_{0010}+\Theta _{0001}\right)  \nonumber \\
\ &\simeq &2\frac{e^{-\frac{N_\tau }\beta }}{1-e^{-\frac 2\beta }}\sum_k\cos
\left( \varphi _k-\varphi _{k+1}\right) +O\left( e^{-2\frac{N_\tau }\beta
}\right)
\end{eqnarray}
are of order $e^{-\frac{N_\tau }\beta }$ . Finalizing we may write 
\begin{equation}
\Theta _{0000}+\Theta _{\left( 1\right) }\simeq \frac 2{1-e^{-\frac 4\beta
}}\left( 1-e^{-\frac{N_\tau }\beta }\left( 1-e^{-\frac 2\beta }\right)
\sum_k\cos \left( \varphi _k-\varphi _{k+1}\right) \right)
\end{equation}
and 
\begin{equation}
1+\frac{\beta _\sigma ^2}2\Xi _2\simeq \left( 1+\frac{\beta _\sigma ^2N_\tau 
}{1-e^{-\frac 4\beta }}\right) \exp \left( -\tilde \beta _{\left( 2\right)
}\sum_k\cos \left( \varphi _k-\varphi _{k+1}\right) \right) +O\left( e^{-2%
\frac{N_\tau }\beta }\right)
\end{equation}
Therefore, for the effective action we get 
\begin{equation}
-S_{eff}\left( \beta ;\beta _\sigma \right) \approx \left( \tilde \beta
-\tilde \beta _{\left( 2\right) }\right) \sum_{\mathbf{x,n}}\cos \left(
\varphi _{\mathbf{x}}-\varphi _{\mathbf{x+n}}\right) +O\left( \beta _\sigma
^4\right) ,
\end{equation}
where 
\begin{equation}
\tilde \beta _{\left( 2\right) }=e^{-\frac{N_\tau }\beta }\left( 1-e^{-\frac
2\beta }\right) \left( 1+\frac{1-e^{-\frac 4\beta }}{\beta _\sigma ^2N_\tau }%
\right) ^{-1}\cdot
\end{equation}

Corrections of higher order in $e^{-\frac{N_\tau }\beta }$ include
'abnormal' terms like $\cos \left( \varphi _{\mathbf{x}}-\varphi _{\mathbf{%
x+n+m}}\right) $\ similar to $\mathbf{\hat I}^2$ in $\left( \ref{sec}\right)
.$ If series in $\beta _\sigma $ converge, corrections in $\beta _\sigma $
are of little importance in the area of $N_\tau >>\beta $.

\section{Appendix C}

We are interested in the limit $N_\tau \rightarrow \infty $\ , therefore, if
there exists such $\beta _{asympt}$\ that for all $\beta >\beta _{asympt}$
then we shall have $\lambda _G^{\left( j\right) }>\lambda _G^{\left(
j_0\right) }=\lambda _{\min }$\ . Therefore, for $T<T_c$\ ($N_\tau >>\lambda
_{\min }$) we may discard all terms\ in $\left( \ref{Aj}\right) $ except
those corresponding to $j=0$ and $j_0$. As it can be shown $\beta
_{asympt}\simeq 5/3$\ $\left( g_{asympt}\simeq 1\right) $\ for $U(1)$\ and $%
\beta _{asympt}\simeq 7/4$\ $\left( g_{asympt}\simeq 3/2\right) $\ for $%
SU(2) $\ gauge group. In all cases, which we consider, $j_0$\ corresponds to
fundamental representation, so we can preserve only two first terms in $%
\left( \ref{Aj}\right) $\ and may write 
\begin{equation}
-\bar S_E\approx N_\tau \ln \Im _0\left( \beta \right) +\exp \left\{ -N_\tau
\lambda _G\left( \beta \right) \right\} \chi \left( \Omega \left( \mathbf{x}%
\right) \right) \chi \left( \Omega \left( \mathbf{x}\right) ^{\dagger
}\right) .
\end{equation}

In such a case instead of $\left( \ref{bess2}\right) $ one may use (\cite
{bateman} 7.13.2$\left( 5\right) $) 
\begin{equation}
I_\nu \left( x\right) =\frac{e^x}{\sqrt{2\pi x}}\sum_{m=0}^{M-1}\left(
-2x\right) ^{-m}\frac{\Gamma \left( \frac 12+\nu +m\right) }{\Gamma \left(
\frac 12+\nu -m\right) \Gamma \left( 1+m\right) }+O\left( x\right) ^{-M}
\label{bess4}
\end{equation}
for more accurate \footnote{%
The asymptotic expansion ($\ref{bess4}$) gives reliable results only for $%
M\lesssim 2\beta e\mathtt{,}$so for $\beta >\beta _{asynpt}$ it would be
legal up to $9$-th term.} computation of beta function, which, taking into
account $\left( \ref{cond}\right) ,$ may be written as 
\begin{equation}
\beta _{CS}\left( g^2\right) \equiv -a_\tau \frac{\partial g}{\partial
a_\tau }=-a_\tau \left( \frac{\partial a_\tau }{\partial g}\right)
^{-1}=-\left( \frac \partial {\partial g}\ln \lambda _G\right) ^{-1}.
\label{bu1}
\end{equation}

In particular for $G=SU(2)$%
\begin{equation}
\beta _{CS}\left( g^2\right) \approx -\frac g2\left( 1-\frac{g^2}8+\left( 
\frac{g^2}8\right) ^2+\frac{17}2\left( \frac{g^2}8\right) ^3+47\left( \frac{%
g^2}8\right) ^4\right) +273\left( \frac{g^2}8\right) ^5.  \label{beta2}
\end{equation}

It is easy to check that if we preserve only two first terms in $\left( \ref
{beta2}\right) $ the beta\ function shows dip at $g_{dip}=\sqrt{8}$%
(non-existent in this case),which, however, is located outside of asymptotic
region. Therefore, to compute $\beta _{CS}\left( g^2\right) $ in given
approximation, we must preserve at least three first terms in $\left( \ref
{beta2}\right) .$ All the next terms are necessary only in the case when
smooth 'link-ups' are needed between $\left( \ref{beta2}\right) $ and $\beta
_{CS}\left( g^2\right) $ computed with the help $\left( \ref{bu1}\right) $
in a strong coupling area $g\gtrsim 1$.


\vspace{1.0in}

\begin{thebibliography}{99}
\bibitem{Creutz}  M. Creutz, '\textit{Quarks, Gluons and Lattices'},
Cambridge: Cambridge Univ. Pr., 1983.

\bibitem{karsch}  F. Karsch, E. Laermann and M. L\"utgemeier,
hep-lat/9411020, '\textit{Three-Dimensional }$SU\left( 3\right) $\textit{\
gauge theory and the Spatial String Tension of the }$\left( 3+1\right) $%
\textit{-Dimensional Finite Temperature }$SU,$ F. Karsch et al., Phys. Lett.%
\textbf{\ B346}, (1995) 94.

\bibitem{hasenfratz}  A. Hasenfratz and P. Hasenfratz, FSU-SCRI-85-2, Ann.
Rev. Nucl. Part. Sci. \textbf{35}(1985)559, '\textit{LATTICE GAUGE THEORIES}%
'.

\bibitem{HHK}  A. Hasenfratz, P. Hasenfratz, U. Heller, and F. Karsch, Phys.
Lett. \textbf{143B}, (1984) 193, '\textit{The }$\beta $\textit{\ - function
of the }$SU(3)$\textit{\ Wilson Action}'

\bibitem{Luscher}  M. Luscher, DESY-97-215, e-Print Archive: hep-ph/9711205,
'\textit{Theoretical advances in a lattice QCD}'. ,Talk given at 18th
International Symposium on Lepton - Photon Interactions (LP 97), Hamburg,
Germany.

\bibitem{BG}  R. Gupta and T. Bhattacharya, Phys. Rev. \textbf{D55} (1997)
7203, T. Bhattacharya and R. Gupta, e-Print Archive: hep-lat 9710095 '%
\textit{Advances in the determination of quark masses}'.

\bibitem{Parisi}  G. Parisi, in: High-Energy Physics --- 1980, XX Int.
Conf., Madison (1980), ed. L. Durand and L. G. Pondrom (American Institute
of Physics, New York, 1981).

\bibitem{LM}  G. P. Lepage and P. Mackenzie, Phys. Rev. \textbf{D48} (1993)
2250.

\bibitem{M}  G. Martinelli et al., Nucl. Phys. \textbf{B445} (1995) 81.

\bibitem{LWW}  M. L\"uscher, P. Weisz and U. Wolff, Nucl. Phys. \textbf{B359}
(1991) 221.

\bibitem{L}  M. L\"uscher et al.~(ALPHA collab.), Nucl. Phys. \textbf{B389}
(1993) 247; \textit{ibid}\/ \textbf{B413} (1994) 481.

\bibitem{D}  G.M. de Divitiis et al.~(ALPHA collab.), Nucl. Phys. \textbf{%
B422} (1994) 382; \textit{ibid}\/ \textbf{B433} (1995) 390; \textit{ibid}\/ 
\textbf{B437} (1995) 447.

\bibitem{J}  K. Jansen et al.~(ALPHA collab.), Phys. Lett. \textbf{B372}
(1996) 275.

\bibitem{C}  S. Capitani et al.~(ALPHA collab.), e-Print Archive: hep-lat
9709125, '\textit{Non-perturbative quark mass renormalization}'.

\bibitem{Biele}  G. Boyd, J. Engels, F. Karsch, E. Laermann, C. Legeland, M.
L\"utgemeier and B. Petersson, BI-TP 96/04, '\textit{Thermodynamics of SU(3)
Lattice Gauge Theory} '.

\bibitem{booth}  S.P.~Booth \textit{et al.}, Phys. Lett. \textbf{B275}
(1992) 424.

\bibitem{patr-seiler}  A. Patrascioiu and E. Seiler, Phys. Rev. Lett. 
\textbf{74} (1995) 1924, Phys. Rev. Lett. \textbf{73} (1994) 3325, Phys.
Rev. Lett. \textbf{76} (1996) 1178, Proceedings of the ICHEP '96, World
Scientific P. Co (1997) p. 1591, '\textit{The problem of asymptotic freedom}'

\bibitem{akemi}  K. Akemi et al., Phys. Rev. Lett. \textbf{71} (1993) 3063.

\bibitem{Uk}  A. Ukawa, Nucl. Phys. B(proc. Suppl.) \textbf{30} 3 (1993).

\bibitem{B}  K.C. Bowler et al., Nucl. Phys. \textbf{B257[FS14]} (1985) 155.

\bibitem{B2}  K.C. Bowler et al., Phys. Lett. \textbf{179B} 375 (1986).

\bibitem{GKPS}  R. Gupta,G.W. Kilcup, A. Patel and S.R. Sharpe, Phys. Lett. 
\textbf{211B} 132 (1988).

\bibitem{H-k}  J. Hoek, Nucl. Phys. \textbf{B339} 732 (1990).

\bibitem{bali1}  G. S. Bali, K. Schilling, J. Fingberg, U. M. Heller and F.
Karsch, , e-Print Archive: hep-lat 9308003,'\textit{Computation of the
Spatial String Tension in High Temperature }$SU\left( 2\right) $\textit{\
Gauge Theory}', G. S. Bali, J. Fingberg, U. M. Heller, F. Karsch and K.
Schilling, Phys. Rev. Lett. \textbf{71} (1993) 3059.

\bibitem{boyd}  G. Boyd, J. Engels, F. Karsch, E. Laermann, C. Legeland, M.
L\"utgemeier, B. Petersson, Nucl. Phys. \textbf{B469} (1996) 419.

\bibitem{MMS}  L.~Maiani, G.~Martinelli, and C.~Sachrajda, Nucl.\ Phys. 
\textbf{B368}, 281 (1992).

\bibitem{petrov}  V. K. Petrov, '\textit{The Spatial string tension in a
finite temperature gluodynamics in a spherical model approximation}',
e-Print Archive: hep-lat 9802031.

\bibitem{green}  F. Green and F. Karsch, Nucl. Phys. \textbf{B238} (1984)
297.

\bibitem{ogilvie}  M. Ogilvie, Phys. Rev. Lett. \textbf{52} (1984)1369,'%
\textit{AN EFFECTIVE SPIN MODEL FOR FINITE TEMPERATURE QCD'}.

\bibitem{billo}  M. Billo, M. Casselle, A. D'Adda, and S. Panzeri, DFTT
69/99, Nordita 96/1p (1996), '\textit{Toward an analytic determination of
the deconfinement temperature in SU(2) L.G.T.}'e-Print Archive: hep-lat
9601020.L.~

\bibitem{aver}  L.~A.~Averchenkova, K.~V.~Petrov, V.~K.~Petrov,
G.~M.~Zinovjev, '\textit{\ Asymmetry parameter role in description of phase
structure of lattice gluodynamics at finite temperature}, Yad. Phys.\textbf{%
\ 60} (1997)1, L.~A.~Averchenkova, K.~V.~Petrov, V.~K.~Petrov,
G.~M.~Zinovjev, '\textit{\ Lattice asymmetry in finite temperature
gluodynamics}, Phys. Rev. \textbf{D56 }, v.11(1997)56.

\bibitem{hasen1}  A.~Hasenfratz and P.~Hasenfratz. Nucl.~Phys.~\textbf{B193}
(1981) 210.

\bibitem{shig}  J.~Shigemitsu, J.~B.~Kogut. Nucl.~Phys.~\textbf{B190} (1981)
3650.

\bibitem{bh-cr}  J.Bhanot and M. Creutz, Phys. Rev. \textbf{D21 }(1980) 2892.

\bibitem{bateman}  H. Batemen and A. Erd\'elyi, '\textit{Higher
Transcendental Functions}' MC Graw-Hill, inc 1953.

\bibitem{BPZ}  O. A. Borisenko, V. K. Petrov, G. M. Zinovjev, TMP, \textbf{77%
}, (1988) 204 and TMP \textbf{80} (1989) 381.

\bibitem{drouffe}  J-M. Drouffe and J-B. Zuber, Phys. Rep. \textbf{102 }%
(1983) 37, '\textit{Strong Coupling and Mean Field Methods in Lattice Gauge
Theories}'.

\bibitem{polyakov}  A. M. Polyakov, Phys. Lett. \textbf{72B }(1978) 477.

\bibitem{susskind}  J. Kogut, L. Susskind, Phys. Rev. \textbf{D9 }(1974)
3501, L. Susskind, Phys. Rev. \textbf{D20}(1979)2610.

\bibitem{gross}  D. J. Gross, R. D. Pisarsky and L. G. Yaffe, Rev. Mod.
Phys. \textbf{53} (1981) 43, '\textit{QCD and instantons at finite
temperature}'

\bibitem{slavnov}  A.A. Slavnov, L.D. Faddeev, 'Introduction into Quantum
Theory of Gauge Fields', Moscow 1978.

\bibitem{svetitsky}  B. Svetitsky and L. G. Yaffe, Nucl. Phys. \textbf{B210}
[FS6] (1982) 423.

\bibitem{ABPZ}  L.~A.~Averchenkova, O. A. Borisenko, V.~K.~Petrov,
G.~M.~Zinovjev, Yad. Phys. \textbf{54 }(1991) 241, 'Higgs Phase in Lattice
QCD Thermodynamics'

\bibitem{karsch2}  F. Karsch, Nucl. Phys. \textbf{B205 }[FS5] (1982) 285, '%
\textit{SU(N) Gauge Theory Couplings on anisotropic Lattices'.}

\bibitem{mack}  G.\ Mack, and V.B. Petkova, Ann.\ Phys.\ \textbf{123} (1979)
442; Ann.\ Phys.\ \textbf{125} (1980) 117; Z.\ Phys.\ \textbf{C12} (1982) 177

\bibitem{yaffe}  L.G. Yaffe, Phys.\ Rev.\ \textbf{D21} (1979) 1574,
L.G.~Yaffe, Phys.\ Rev.\ \textbf{D21}, 1574, (1980).

\bibitem{T}  E.T.~Tomboulis, Phys.\ Rev.\ \textbf{D23}, 2371, (1981); in
''Proceedings of the Brown Workshop on Nonperturbative Studies in QCD'', A.
Jevicki and C-I. Tan, eds., (1981), E.T.~Tomboulis, Phys.\ Lett.\ \textbf{%
B303}, 103, (1993); Nucl.\ Phys.\ \textbf{B34} (1994) 192, (Proc.\ Suppl.);
T.G.~Kov\'acs and E.T.~Tomboulis, Nucl.\ Phys.\ \textbf{B53} (1997) 509,
(Proc.\ Suppl.); \textbf{\ }e-Print Archive: hep-lat/9709042.

\bibitem{sim}  E.L. Gubankova and Yu.A. Simonov,\textbf{\ }e-Print Archive:
hep-lat 9508206 '\textit{MAGNETIC CONFINEMENT AND SCREENING MASSES}'.

\bibitem{pier}  S. W.~Pierson, e-Print Archive: cond-mat 9711040, '\textit{A
Review of the Real Space Renormalization Group Analysis of the Two-
Dimensional Coulomb Gas, the Kosterlitz-Thouless-Berezinskii Transition, and
Extensions to a Layered Vortex Gas}'.

\bibitem{BC-HMP}  P. Butera and M. Comi, \textbf{Preprint IFUM 429/FTm July
1992, }e-Print Archive: hep-lat/9209011\textbf{, }and M. Hasenbusch, M.
Marcu and K. Pinn, DESY-92-010, Jan 1992, in *Tsukuba 1991, Proceedings,
Lattice 91, e-Print Archive: hep-lat 9207019 '\textit{High} \textit{%
Precision Verification of the Kosterlitz Thouless Scenario}'.

\bibitem{alonso}  J.L. Alonso, J.M. Carmona, J. Clemente Callardo, L.A.
Fernanandez, D. Iniguez, A. Tarancon, C.L. Ullod, e-Print Archive: hep-lat
9503016.

\bibitem{alv-ol}  O. Alvarez, Phys. Rev. \textbf{D24} (1981) 440 and P.
Olesen, Phys. Lett. \textbf{160B} (1985) 408.

\bibitem{alles}  B.\ All\'es, D.S.\ Henty, H.\ Panagopoulos, C.\ Parrinello,
C.\ Pittori, D.G.\ Richards, hep-lat 9605033, '$\alpha _s(\mu )$ \textit{\
from the Non-perturbatively Renormalized Lattice Three-gluon Vertex}'

\bibitem{kogut}  J.B. Kogut, Rev. Mod. Phys., \textbf{55}, (1983), 800.

\bibitem{parisi}  G. Parisi, in \textit{High Energy Physics - 1980},
Proceedings of the XXth International Conference, Madison, Wisconsin, 1980
(AIP, New York, 1981), p.1531.

\bibitem{engels}  J. Engels., F. Karsch, and K. Redlich, BI-TP 94/30,
(1997), '\textit{Scaling Properties of the Energy Density in SU(2) Lattice
Gauge Theory}'

\bibitem{agostini}  V. Agostini, G. Carlino, M. Cselle and M. Hasenbusch,
DFTT 23/96, HUB-EP-96/24, '\textit{The Spectrum of the }$2+1$\textit{\
Dimensional Gauge Ising model}'

\bibitem{guth}  A.~Ukawa, P.~Windey, A.H.~Guth. Phys.Rev. \textbf{D21}
(1980) 1013

\bibitem{Billoire}  A. Billoire. Phys. Lett. \textbf{104 B},472 (1981)

\bibitem{L-W et al}  M. L\"uscher, P. Weisz and U. Wolff, Nucl. Phys. 
\textbf{B359} (1991) 221, M. L\"uscher et al.~(ALPHA collab.), Nucl. Phys. 
\textbf{B389} (1993) 247; Nucl. Phys.\/\textbf{\ B413} (1994) 481, G.M. de
Divitiis et al.~(ALPHA collab.), Nucl. Phys. \textbf{B422} (1994) 382;
\/Nucl. Phys. \textbf{B433} (1995) 390; \/Nucl. Phys. \textbf{B437} (1995)
447, K. Jansen et al.~(ALPHA collab.), Phys. Lett. \textbf{B372} (1996) 275,
S. Capitani et al, (ALPHA collab.), hep-lat/9709125,~'\textit{%
Non-perturbative quark mass renormalization'}.
\end{thebibliography}
\end{document}